\documentclass[12pt,fleqn,a4paper]{article}
\usepackage{graphicx}
\usepackage{latexsym}
\usepackage[small,bf]{caption2}
\usepackage{amsfonts}
\usepackage{amssymb}
\usepackage{amsmath}
\usepackage{theorem}
\usepackage{times}
\usepackage[numbers,sort&compress]{natbib}
\newcommand{\bea}{\begin{eqnarray}}
\newcommand{\eea}{\end{eqnarray}}
\newcommand{\be}{\begin{equation}}
\newcommand{\ee}{\end{equation}}
\newcommand{\bef}{\begin{figure}}
\newcommand{\enf}{\end{figure}}
\newcommand{\ball}{\begin{array}{ll}}
\newcommand{\bacl}{\begin{array}{cl}}
\newcommand{\bal}{\begin{array}{l}}
\newcommand{\bac}{\begin{array}{c}}
\newcommand{\ea}{\end{array}}
\newcommand{\feta}{{\boldsymbol{\eta}}}

\newcommand{\fzeta}{{\boldsymbol{\zeta}}}

\newcommand{\mubo}{{\boldsymbol{\mu}}}
\newcommand{\rhobo}{{\boldsymbol{\rho}}}
\newcommand{\thetabo}{{\boldsymbol{\theta}}}

\newcommand{\psibo}{{\boldsymbol{\phi}}}

\newcommand{\Sigmabo}{{\boldsymbol{\Sigma}}}
\newcommand{\grad}{{\boldsymbol{\nabla}}}

\newcommand{\N}{{\mathbb{N}}}
\newcommand{\R}{{\mathbb{R}}}
\newcommand{\ganz}{{\mathbb{Z}}}

\newcommand{\ord}{{O}}
\newcommand{\abo}{{\mathbf{a}}}

\newcommand{\xbo}{{\mathbf{x}}}

\newcommand{\nbo}{{\mathbf{n}}}
\newcommand{\kbo}{{\mathbf{k}}}
\newcommand{\ebo}{{\mathbf{e}}}
\newcommand{\Rbo}{{\mathbf{R}}}

\newcommand{\Mbo}{{\mathbf{M}}}
\newcommand{\Nbo}{{\mathbf{N}}}
\newcommand{\Di}{{\mathrm{int}\, D}}
\newcommand{\dom}{{\rm dom\,}}
\newcommand{\lev}{{\rm lev}}

\numberwithin{equation}{section}
\theoremstyle{plain}
\newtheorem{theorem}{Theorem}[section]
\newtheorem{lemma}[theorem]{Lemma}
\newtheorem{corollary}[theorem]{Corollary}
\newtheorem{proposition}[theorem]{Proposition}
\title{Equivalence of ensembles for two-species zero-range invariant measures}
\author{Stefan Grosskinsky\thanks{Mathematics Institute, Zeeman Building, University of Warwick, Coventry CV4 7AL, UK.\newline email: S.W.Grosskinsky@warwick.ac.uk}\\ \small University of Warwick}

\begin{document}
\maketitle

\begin{abstract}
We study the equivalence of ensembles for stationary measures of interacting particle systems with two conserved quantities and unbounded local state space. The main motivation is a condensation transition in the zero-range process which has recently attracted attention. Establishing the equivalence of ensembles via convergence in specific relative entropy, we derive the phase diagram for the condensation transition, which can be understood in terms of the domain of grand-canonical measures. Of particular interest, also from a mathematical point of view, are the convergence properties of the Gibbs free energy on the boundary of that domain, involving large deviations and multivariate local limit theorems of subexponential distributions.
\end{abstract}

\textbf{keywords.} zero-range process; equivalence of ensembles; condensation transition; relative entropy


\section{Introduction}
Zero-range processes are interacting particle systems with no restriction on the number of particles per site, i.e.\ with unbounded local state space. The jump rate of each particle depends only on the number of particles at its departure site which leads to a simple product structure of the stationary measure \cite{spitzer70,andjel82}. These processes have recently attained much attention in the theoretical physics literature (see \cite{evansetal05} and references therein) since they exhibit a condensation transition under certain conditions on the jump rates \cite{evans00}. If the particle density exceeds a critical value $\rho_c$, the system phase separates into a homogeneous background with density $\rho_c$ and a condensate, where the excess particles accumulate. First rigorous results on a single species system \cite{stefan} show that this phase transition can be understood mathematically in the context of the equivalence of ensembles. This is a classical problem of mathematical statistical mechanics \cite{ruelle69} which arises naturally in the context of studying stationary measures of interacting particle systems with conserved quantities, such as energy or the number of particles. In general, interacting particle systems with several conservation laws are currently of particular interest in non-equilibrium statistical mechanics, since they show a very rich critical behaviour (see \cite{schuetz03} and references therein). There are not many general results for such systems, for example for zero-range processes with more than one particle species there exist only non-rigorous case studies so far \cite{evansetal03}. The motivation of this paper is to understand the condensation transition in such multi-species processes on the rigorous level of the equivalence of ensembles.

For simplicity of presentation we focus on systems with two conserved quantities, which we interpret as the number of particles in a two-species system. The local state space for each species is $\N =\{ 0,1,\ldots \}$, i.e.\ the numbers of particles on each lattice site are unrestricted. We require that the process has a stationary measure of product form. Due to the conservation law this induces a family of stationary measures $\pi_{L,\Nbo}$ with fixed particle numbers $\Nbo\in\N^2$ on a lattice of size $L$, the canonical ensemble. Another standard family is the grand-canonical ensemble $\nu_\mubo^L$, where the numbers of particles are random variables. The densities $\rhobo\in (0,\infty )^2$, the expected numbers of particles per site, are controlled by conjugate parameters $\mubo\in\R^2$, the chemical potentials. In our case $\nu_\mubo^L$ is a product measure and is also defined for $L\to\infty$, where we write $\nu_\mubo$. Let $D_\mu \subset\R^2$ denote the maximal domain, such that $\nu_\mubo^1$ is normalizable with finite first moment.

In the thermodynamic limit $\Nbo_L /L\to\rhobo$ as $L\to\infty$ with densities $\rhobo$, one expects that
\be\label{eoes}
\exists\,\mubo (\rhobo )\in D_\mu \ :\ \pi_{L,\Nbo_L } \to \nu_{\mubo (\rhobo )} \quad\mbox{as }L\to\infty\ .
\ee
The question of the equivalence of ensembles is for which values of $\rhobo$ and in what sense (\ref{eoes}) holds, and how it has to be modified in the presence of phase separation. The main results of this paper are:
\begin{itemize}
\item[1.] We establish the equivalence of ensembles (\ref{eoes}) for all $\rhobo\in (0,\infty )^2$ under mild regularity assumptions on the stationary product measure.
\end{itemize}
In the proof we use specific relative entropy (or relative information gain), which is based on results from information theory \cite{csiszar84} and was previously applied to study large deviations and the equivalence of ensembles for Gibbsian random fields \cite{georgii93}, marked point processes \cite{georgiietal93} and weakly dependent measures \cite{lewisetal95}. A common feature of these models is a bounded Hamiltonian, which corresponds to $D_\mu =\R^2$ in the above setting. In this case, phase separation is a consequence of long-range correlations, leading to non-differentiability of the Gibbs free energy (or non-convex canonical entropy) and a first order transition \cite{ellisetal00,touchetteetal04}. In our case there are no spatial correlations, but typically $D_\mu \subsetneq\R^2$ due to the unbounded local state space, and condensation is a result of large deviation properties of $\nu_\mubo^1$ on the boundary of $D_\mu$, where it turns out to be subexponential.
\begin{itemize}
\item[2.] We show how the phase diagram for the condensation transition can be derived solely from the shape of $D_\mu$, and explain its relation to the mode of convergence in (\ref{eoes}).
\end{itemize}
The transition is continuous and is characterized by convergence properties of the Gibbs free energy on the boundary of $D_\mu$. In the classification of \cite{touchetteetal04} this corresponds to the case of partial equivalence of ensembles.

Our results can be directly generalized to any number of particle species with arbitrary discrete local state spaces. We choose to work in a more specific setting for the simplicity of presentation, since it covers the basic novelties of the paper. From the point of view of non-equilibrium statistical mechanics these are the first rigorous results on the condensation transition in a system with several conservation laws. From a mathematical point of view, we adapt the theory of the equivalence of ensembles to study phase separation in systems with unbounded Hamiltonians. Even in the basic case of stationary product measures the different mathematical origin of the condensation transition leads to interesting new aspects. Our equivalence result involves a sharp condition on the number of particles and is valid on the (non-empty) boundary of $D_\mu$. The analysis requires results on large deviations \cite{baltrunasetal02,vinogradov} and multi-dimensional local limit theorems of subexponential distributions \cite{rvaceva54,mukhin91}, as well as convergence properties of multivariate power series similar to \cite{pemantleetal05}. In contrast to a previous study for single-species processes \cite{stefan}, the present paper provides a complete picture of the mechanism of condensation in a much more general context.

Precise definitions of the ensembles and basic properties are given in the next section. The main results are given in Section 3, including the equivalence of ensembles and the construction of the phase diagram for the condensation transition. For completeness, we also include some remarks on fluctuations and the spatial extension of the condensate (cf. \cite{jeonetal00,stefan}). Proofs are given in Section 4. Since the main results apply for ensembles of measures in a general context, the paper up to this point is formulated without reference to zero-range processes, which are, however, the main motivation for this study. In Section 5 we explain why these processes provide a natural class of particle systems for the measures considered in the first sections, and illustrate the results on the phase diagram by several examples. Some results from convex analysis needed in the proof of the equivalence of ensembles are summarized in the appendix.\\

\section{Preliminaries}

  \subsection{Canonical and grand-canonical measures}
Consider $L$ independent identically distributed random vectors
\be
\feta (x)=\big(\eta_1 (x),\eta_2 (x)\big)\in\N^2 \ ,\quad x\in\Lambda_L \ ,
\ee
with some discrete index set $\Lambda_L$ of size $|\Lambda_L |=L$. The state space $X_L =(\N^2 )^{\Lambda_L}$ is a measure space with $\sigma$-algebra induced by the product topology and the (a-priori) measure
\be\label{weight}
w^L (\feta )=\prod_{x\in\Lambda_L} w\big(\feta (x)\big)\in (0,\infty )\quad\mbox{for}\quad\feta =\big(\feta (x )\big)_{x\in\Lambda_L} \ .
\ee
This should be positive but not necessarily normalized, i.e. $w:\N^2 \to (0,\infty )$ is arbitrary. Since $X_L$ is discrete, we simplify notation here and in the following by using the same symbols for a measure and its mass function, i.e. $w^L (\feta )=w^L \big(\{\feta\}\big)$.

We interpret the index set $\Lambda_L$ as a lattice of size $L$ and $\feta \in X_L$ as particle configurations of a two-species particle system. We do not specify the geometry of the lattice, boundary conditions or dynamics of this process, they should be such that $w^L$ is a stationary weight, i.e.\ up to normalization, $w^L$ is a stationary distribution of the process. Generic particle systems with this property are zero-range processes discussed in Section 5. Apart from stationarity of $w^L$, the only other requirement on the particle system is that the numbers of particles
\be
\Sigmabo_L (\feta )=\big(\Sigma^1_L ,\Sigma^2_L \big) (\feta ):=\sum_{x\in\Lambda_L}\feta (x)\in\N^2
\ee
are conserved quantities for each species and that there are no other conservation laws. Then there exists a family of stationary probability measures $h(\Sigmabo_L )\, w^L$ which are absolutely continuous with respect to $w^L$, where the Radon-Nikodym derivative depends only on the conserved quantities $\Sigmabo_L$ and can be written as a function $h:\N^2 \to [0,\infty )$.

The set of all stationary measures of the particle system is convex and the extremal measures are given by choosing $h(\Sigmabo_L )\propto\delta_{\Sigmabo_L ,\Nbo}$, i.e.\ proportional to the Kronecker delta, fixing the number of particles to $\Nbo =(N_1 ,N_2 )\in\N^2$. The family
\be\label{ce}
\pi_{L,\Nbo} (\feta )=\frac{1}{Z_{L,\Nbo }}\prod_{x\in\Lambda_L}
w\big(\feta (x)\big)\,\delta_{\Sigmabo_L (\feta
  ),\Nbo}\ ,\quad\Nbo\in\N^2
\ee
is the \textit{canonical ensemble} and the measures concentrate on finite subsets
\be
X_{L,\Nbo } =\big\{\feta\,\big|\, \Sigmabo_L (\feta )=\Nbo\big\}\subsetneq X_L
\ee
of configurations with fixed particle numbers. The canonical partition function is $Z_{L,\Nbo} =w^L (X_{L,\Nbo})\in (0,\infty )$, since $\pi_{L,\Nbo} =w^L \big(\, .\,\big|\, \{\Sigmabo_L =\Nbo\} \big)$ can be written as a conditional measure. By assumption, for each fixed $L\geq 1$ and $\Nbo\in\N^2$ the particle system is irreducible on $X_{L,\Nbo }$ and $\pi_{L,\Nbo}$ is the unique stationary measure. All other stationary measures on $X_L$ are convex combinations of canonical measures.

Another generic choice is $g(\Sigmabo_L )\propto e^{\mubo\cdot\Sigmabo_L}$ with parameters $\mubo =(\mu_1 ,\mu_2 )\in\R^2$ called chemical potentials, defining the \textit{grand-canonical measures}
\be\label{gce}
\nu^L_\mubo (\feta )=\frac{1}{z(\mubo )^L}\,\prod_{x\in\Lambda_L} w\big(\feta (x)\big)\, e^{\mubo\cdot\feta (x)}\ .
\ee
Each $\nu_\mubo^L$ is supported on $X_L$, i.e.\ $\Sigmabo_L$ is a random variable and the expected value is fixed by the chemical potentials $\mubo$, as is discussed below. These measures are particularly convenient since they are of product form. The normalizing (single site) partition function
\be\label{part}
z(\mubo )=\sum_{\kbo \in\N^2} w(\kbo )\, e^{\mubo\cdot\kbo} 
\ee
is an infinite sum, as opposed to models with bounded local state space, such as $\{ 0,1\}$ for lattice gases or $\{ -1,1\}$ for spin systems. For such systems, $z(\mubo )$ is defined for all $\mubo\in\R^2$, whereas in our case the domain of definition of $z$ will play a crucial role.

  \subsection{Properties of grand-canonical measures}
We define
\be\label{dmu}
D_\mu =\Big\{ \mubo\in\R^2\, \Big|\, \sum_{\kbo\in\N^2} k_i \, w(\kbo )\, e^{\mubo\cdot\kbo} <\infty\mbox{ for }i=1,2\Big\}\ .
\ee
This implies that for all $\mubo\in D_\mu$, $z(\mubo )<\infty$ and the product measure $\nu_\mubo$ is well defined. Moreover, on $D_\mu$ the marginal $\nu_\mubo^1$ has finite first moments, which are interpreted as particle densities and given by
\be\label{rho2}
\Rbo =(R_1 ,R_2 ):D_\mu \to (0,\infty )^2\ ,\quad\mbox{where}\quad R_i (\mubo )=\langle\eta_i \rangle_{\nu^1_\mubo}\ ,\ i=1,2\ .
\ee
Here and in the following we write $\langle ..\rangle_{\nu}$ for the expected value with respect to measure $\nu$. Note that $R_i (\mubo ) =\big\langle\eta_i (x) \big\rangle_{\nu^L_\mubo}$ independently of the lattice site $x\in\Lambda_L$, and that $D_\mu$ as defined in (\ref{dmu}) is the maximal domain of definition of $\Rbo$, i.e. $D_\mu =\dom\Rbo$. We denote by
\be
D_\rho =\Rbo \big( D_\mu \big)\subset (0,\infty )^2
\ee
the range of $\Rbo$, which characterizes the set of all densities accessible by the grand-canonical ensemble. 

In the following we assume that $w$ is exponentially bounded, i.e.
\be\label{cond1}
\exists\,\xi\in (0,\infty)\ \forall\,\kbo\in\N^2 \ :\ w(\kbo )\leq\xi^{|\kbo |}\ ,
\ee
where we write $|\kbo |=\|\kbo\|_2 =\big (k_1^2 +k_2^2 \big)^{1/2}$. For convenience we further assume that the single site mass function $w$ is actually defined on $[0,\infty )^2$ with
\be\label{cond1c}
w\in C^1 \big( [0,\infty )^2 ,(0,\infty )\big)\ ,
\ee
which imposes no restriction on the relevant values $w(\kbo ), \kbo\in\N^2$.

\begin{lemma}\label{theo21}
$D_\mu \neq\emptyset$ (and thus $D_\rho \neq\emptyset$) if and only if (\ref{cond1}) is fulfilled. In this case $D_\mu$ is convex and complete, i.e.
\be\label{deltadef}
\Delta (\mubo^* ):=\{\mubo\, |\,\mu_i \leq\mu_i^* ,\, i=1,2\}\subset D_\mu \quad\mbox{whenever}\quad\mubo^* \in D_\mu \ .
\ee
Either $D_\mu =\R^2$ or the boundary can be characterized in the rotated variables $\tilde\mu_1 =\mu_1 -\mu_2$ and $\tilde\mu_2 =\mu_1 +\mu_2$ by\quad $\partial D_\mu =\big\{ (\tilde\mu_1 ,\tilde\mu_2 (\tilde\mu_1 ))\big|\,\tilde\mu_1 \in\R\big\}$. Here $\tilde\mu_2 :\R\to\R$ is continuous and piecewise differentiable, with
\be\label{ddmu}
\tilde\mu_2 (\tilde\mu_1 )=-\limsup_{|\kbo |\to\infty} \Big( 2\log w(\kbo )+\tilde\mu_1 (k_1 -k_2 )\Big)\big/ (k_1 +k_2)\ .
\ee
All the above properties also hold for $\dom z$, the maximal domain of definition of $z$. We have $D_\mu \subset\dom z$ and ${\rm int\,} D_\mu ={\rm int\,}\dom z$ for the interior, so both sets are equal or differ only on the boundary.
\end{lemma}

Since the grand-canonical measures are product measures the pressure is given by
\be
p(\mubo )=\lim_{L\to\infty}\frac1L\log z(\mubo )^L =\log z(\mubo )\ ,
\ee
which is the analogue of the Gibbs free energy. For all $\mubo\in D_\mu$ the density (\ref{rho2}) can be written as
\be\label{rhodp}
R_i (\mubo )=\partial_{\mu_i} p(\mubo )\ ,\quad i=1,2\ .
\ee
The derivatives are defined one-sided on $\partial D_\mu \cap D_\mu$, which is possible due to completeness of $D_\mu$ and the following lemma.

\begin{lemma}\label{theo21b}
The single site marginal $\nu^1_\mubo$ has some finite exponential moments if and only if $\mubo\in\Di_\mu$. Moreover, $p\in C^\infty (\Di_\mu ,\R )$, $p\in C^1 (D_\mu ,\R )$ and $p$ is strictly convex on $D_\mu$.\\
$p$ and $\Rbo$ can be extended continuously to
\be
\partial^{1,-\infty} D_\mu =\big\{ (-\infty ,\mu_2 )\,\big|\,\exists \mu_1 \in\R\, :(\mu_1 ,\mu_2 )\in D_\mu \big\}\ ,
\ee
i.e. limits exist and are given by
\be
p(-\infty ,\mu_2 )=\sum_{k_2 =0}^\infty w(0,k_2 )\, e^{\mu_2 k_2}\quad\mbox{and}\quad \Rbo (-\infty ,\mu_2 )={0\choose \partial_{\mu_2} p(-\infty ,\mu_2 )}\ .
\ee
An analogous result holds for $\partial^{2,-\infty} D_\mu$.\\
For $i=1,2$, if $D_\rho$ is bounded in $\rho_i$, i.e. for all $\rhobo\in D_\rho$, $\rho_i \leq C$ for some $C\geq 0$, then $D_\mu$ is bounded in $\mu_i$.
\end{lemma}

Since $p$ is strictly convex, $\Rbo$ is invertible on $D_\mu$ due to (\ref{rhodp}) and we denote the inverse by $\Mbo :D_\rho \to D_\mu$. The entropy density $s:(0,\infty )^2 \to\R$ of the grand-canonical measure (\ref{gce}) is the convex conjugate of the pressure given by the Legendre transform (cf. (\ref{legend}))
\be\label{entropy}
s(\rhobo )=p^* (\rhobo )=\sup_{\mubo\in D_\mu}\big(\rhobo\cdot\mubo - p(\mubo )\big)\ .
\ee
Thus $s$, also known as the large deviation rate function, is strictly convex on $D_\rho$ and convex on $(0,\infty )^2$. For $\rhobo\in \Di_\rho$ it is easy to see that $\rhobo\cdot\mubo - p(\mubo )$ has a local maximum at $\Mbo (\rhobo )$ and thus
\be
s(\rhobo )=\rhobo\cdot\Mbo (\rhobo )-p\big(\Mbo (\rhobo )\big)\quad\mbox{and}\quad M_i (\rhobo )=\partial_{\rho_i} s(\rhobo )\ ,\ i=1,2\ .
\ee
Using convexity of $D_\mu$ and $p(\mubo )$ we can show that there exists a unique maximizer of the right hand side of (\ref{entropy}), also for $\rhobo\not\in \Di_\rho$. This is the main result of this preliminary section.

\begin{proposition}\label{theo22}
For every $\rhobo\in (0,\infty )^2$ there exists a unique maximizer $\overline{\Mbo}(\rhobo )\in D_\mu$ of the right hand side of (\ref{entropy}), such that
\be
s(\rhobo ) =\rhobo\cdot \overline{\Mbo} (\rhobo )-p\big(\overline{\Mbo} (\rhobo )\big)\ .
\ee
$\overline\Mbo\in C\big( (0,\infty )^2 ,\R\big)$ and we have $\overline{\Mbo} (\rhobo )=\Mbo (\rhobo )$ for $\rhobo\in D_\rho$ and $\overline{\Mbo} (\rhobo )\in\partial D_\mu \cap D_\mu$ for $\rhobo\not\in D_\rho$.\\
In particular, $D_\rho$ is closed in $(0,\infty )^2$ and $\partial D_\rho =\Rbo (\partial D_\mu \cap D_\mu )$, where $\partial D_\rho$ denotes the relative boundary of $D_\rho$ in $(0,\infty )^2$.\\
\end{proposition}

\section{Main Results}
  \subsection{Equivalence of ensembles}
Consider a sequence of canonical measures $\pi_{L,\Nbo_L}$ in the thermodynamic limit, i.e.
\be\label{theol}
\Nbo_L /L\to\rhobo\quad\mbox{as }L\to\infty\quad\mbox{with density $\rhobo\in (0,\infty )^2$.}
\ee
In the following we study the question if the sequence $\pi_{L,\Nbo_L}$ converges to a grand-canonical product measure, and if yes, what is the mode of convergence. To quantify the distance between the measures we use the specific relative entropy
\bea
h_{L,\Nbo } (\mubo )&=&\frac1L\, H\big(\pi_{L,\Nbo} \big|\nu_\mubo^L \big)\ ,\qquad\mbox{where}\nonumber\\
H\big(\pi_{L,\Nbo} \big|\nu_\mubo^L \big) &=&\sum_{\feta\in X_L} \pi_{L,\Nbo}(\feta )\,\log\frac{\pi_{L,\Nbo} (\feta )}{\nu_\mubo^L (\feta )}
\eea
is the usual relative entropy, since $\pi_{L,\Nbo}$ is absolutely continuous with respect to $\nu_\mubo^L$. Using the relations
\bea
\nu_\mubo^L (\feta )\, z(\mubo )^L &=&w^L (\feta )\, e^{\mubo\cdot\Nbo}\qquad\mbox{for all }\feta\in X_{L,\Nbo }\quad\mbox{and}\nonumber\\
\nu_\mubo^L \big(\{\Sigmabo_L =\Nbo\}\big)\, z(\mubo )^L &=&Z_{L,\Nbo} \, e^{\mubo\cdot\Nbo}\ ,
\eea
which are easily derived from (\ref{ce}) and (\ref{gce}), we can write
\be\label{forms}
h_{L,\Nbo } (\mubo )=-\frac1L\,\log\nu_\mubo^L \big(\{\Sigmabo_L =\Nbo\}\big) = p(\mubo )-\frac{\mubo\cdot\Nbo}{L} -\frac1L\log Z_{L,\Nbo }\ ,
\ee
for all $L\geq 1$, $\Nbo\in\N^2$ and $\mubo\in D_\mu$.

The second part of (\ref{forms}) suggests that $\overline\Mbo (\rhobo )$ of Proposition \ref{theo22} is the right chemical potential to minimize $h_{L,\Nbo_L }$ in the thermodynamic limit (\ref{theol}). This is the content of the next theorem for which we need a further regularity assumption on the exponential tail of $w$, in addition to (\ref{cond1}) and (\ref{cond1c}). A convenient sufficient condition is that for all $\phi\in [0,\pi /2]$ the limit in the radial direction $\ebo_\phi$
\be\label{cond2}
\lim_{r \to\infty}\frac1{r}\,\log w( r\,\ebo_\phi )\in\R\quad\mbox{exists}\ ,
\ee
and is a continuous function of $\phi$. This can be relaxed considerably as is discussed after the proof in Section 4.3. (\ref{cond2}) holds for example if $w$ is convex, or if $w=w_1 +w_2$ where $w_1$ is convex and $w_2$ has bounded derivative.

\begin{theorem}\label{theo31}
Assume (\ref{cond1}), (\ref{cond1c}) and (\ref{cond2}). Then for each particle density $\rhobo\in (0,\infty )^2$ and every sequence $\Nbo_L$ as in (\ref{theol})
\bea\label{eoe}
\lim_{L\to\infty} h_{L,\Nbo_L } \big(\overline\Mbo (\rhobo )\big) =0\ .\\
\nonumber
\eea
\end{theorem}
From this result one can immediately deduce two standard formulations of the equivalence of ensembles, on the level of measures and on the level of thermodynamic functions. To formulate the first version we have to define all canonical and all grand-canonical measures on a common state space $X=\N^\Lambda$, where $\Lambda$ is the (infinite) limit lattice of an appropriate sequence $(\Lambda_L )_{L=1,2,..}$. The precise construction is deferred to Appendix B, since it is only necessary to formulate (\ref{corr1}) and has no further importance for our results.

\begin{corollary}\label{theo32}
For each $\rhobo\in (0,\infty )^2$ we have
\be\label{corr1}
\langle f\rangle_{\pi_{L,\Nbo_L}}\to\langle f\rangle_{\nu_{\overline\Mbo (\rhobo )}}\quad\mbox{as }L\to\infty\ ,
\ee
for all cylinder test functions $f\in C(X,\R )$ with $\langle e^{\epsilon f}\rangle_{\nu_{\overline\Mbo (\rhobo )}}<\infty$ for some $\epsilon >0$. In particular, this includes all bounded $f\in C_b (X,\R )$, which is equivalent to convergence in distribution. Moreover,
\be\label{corr2}
\lim_{L\to\infty} \frac1L\,\log Z_{L,\Nbo_L} =-s(\rhobo )\ .
\ee
\end{corollary}
For $\rhobo\in\Di_\rho$, $\nu_{\overline\Mbo (\rhobo )}$ has some finite exponential moments by Lemma \ref{theo21}, so in particular the corollary implies convergence of the local densities $f(\feta )=\eta_i (x)$. We note that for a single species with $\rhobo\in \Di_\rho$ convergence is shown even for $L^2$ test functions in \cite{kipnislandim}, Appendix 2.1. The proof given there relies on rather involved estimates on the rate of convergence in the local limit theorem, whereas the proof via relative entropy is much simpler (see section 4.3). Moreover, our result covers several particle species and can be generalized to $\rhobo\not\in D_\rho$, which is the main point of this paper. In this case the nature of the convergence changes and (\ref{corr1}) is violated for $f(\feta )=\eta_i (x)$ for at least one species $i$, as will become clear in the next subsection. This difference in the mode of convergence is a result of the unbounded local state space and is a signature of the condensation transition.

For systems with bounded local state space (\ref{eoe}) implies convergence for all cylinder test functions $f\in C(X,\R )$. But in case of $\rhobo\not\in D_\rho$ the limiting measure would be a mixture of grand-canonical measures, corresponding to coexisting domains with different distributions for large finite systems (see e.g.\ \cite{ellisetal00}). This phenomenon is called phase separation. In analogy to this classical case we interpret our limit result in the following way: For $\rhobo\not\in D_\rho$ the system phase separates into a \textit{(homogeneous) background phase} with product measure $\nu_{\overline\Mbo (\rhobo )}$ given by Theorem \ref{theo31}, and a \textit{condensate} or \textit{condensed phase} which contains the excess particles. According to (\ref{corr1}), the condensate cannot be tested by cylinder functions in the infinite system, its existence is only a consequence of the conservation law (in contrast to classical phase separation). The interpretation for large finite systems is that the volume fraction covered by the condensate domain vanishes as $L\to\infty$. In fact, this domain typically concentrates only on a single lattice site, which is proved under additional assumptions in Section 3.3.

Due to the conservation laws, the phase space of the particle system is $(0,\infty )^2$, the set of densities $\rhobo$. We say that the particle system exhibits a condensation transition, if $D_\rho \subsetneq (0,\infty )^2$. As order parameter of the phase transition we choose the mapping
\be\label{rhoc}
\Rbo_c :(0,\infty )^2 \to D_\rho\ ,\quad\mbox{with}\quad \Rbo_c (\rhobo ):=\Rbo\big(\overline{\Mbo}(\rhobo )\big)\ .
\ee
According to the above interpretation, $\Rbo_c (\rhobo )$ describes the density of the background phase in a system with global density $\rhobo$. Note that by Proposition \ref{theo22}
\be
\Rbo_c (\rhobo )\left\{\bacl =\rhobo\ &,\mbox{ if }\rhobo\in D_\rho \\ \in \partial D_\rho\ &,\mbox{ if }\rhobo\not\in D_\rho\ea\right.\ ,
\ee
so $\Rbo_c (D_\rho )=D_\rho$ and $\Rbo_c$ is a projection from $(0,\infty )^2$ onto $D_\rho$. By Lemma \ref{theo21b} and Proposition \ref{theo22}, $\Rbo_c \in C\big( (0,\infty )^2 ,D_\rho\big)$ so the transition is continuous (second order), which is directly related to the fact that $p\in C^1 (D_\mu ,\R )$. This is in contrast to systems with bounded local state space, where we would have $D_\mu =\R^2$ and non-differentiability of $p$ would lead to a first order phase transition with discontinuous order parameter \cite{ellisetal00,touchetteetal04}.

  \subsection{Phase diagram}
In this section we apply standard results from convex analysis, which are summarized in Appendix A, to characterize the phase diagram of the system. By Proposition \ref{theo22}, $\partial D_\rho =\Rbo (\partial D_\mu \cap D_\mu )$, and thus condensation occurs if and only if $\partial D_\mu \cap D_\mu \neq\emptyset$.

\begin{theorem}\label{theo33}
For every $\rhobo_c \in\partial D_\rho$ with $\mubo =\Mbo (\rhobo_c )$, the preimage $\Rbo_c^{-1} (\rhobo_c )$ is given by the subgradient $\delta p(\mubo )$ as defined in (\ref{subgr}). Moreover,
\be\label{subdiff}
\delta p(\mubo ) =\left\{\bacl\big\{\rhobo_c +\lambda\,\nbo_\mubo \,\big|\,\lambda\geq 0\big\}\ &,\ \partial D_\mu\mbox{ diff'able in }\mubo\\
\big\{\rhobo_c {+}\lambda^+ \,\nbo^+_\mubo {+}\lambda^- \,\nbo^-_\mubo\,\big|\,\lambda^+ ,\lambda^- \geq 0\big\}\ &,\ \mbox{ otherwise}\ea\right. .
\ee
$\nbo_\mubo$ denotes the normal vector to $\partial D_\mu$ in $\mubo$ and $\nbo^+_\mubo$, $\nbo^-_\mubo$ the two limiting normal vectors, in case $\partial D_\mu$ is not differentiable in $\mubo$.
\end{theorem}
Note that by convexity of $D_\mu$, $\nbo_\mubo^+$ and $\nbo_\mubo^-$ are well defined as the extremal normal directions to the set of supporting hyperplanes in $\mubo$. In case that the points of non-differentiability of $\partial D_\mu$ (see Lemma \ref{theo21}) accumulate in $\mubo$, $\nbo_\mubo^+ =\nbo_\mubo^-$ is also possible. In the following we use Theorem \ref{theo33} to construct the phase diagram and its properties.

By definition (\ref{rhoc}), the preimage
\be
\Rbo_c^{-1} (\rhobo_c )=\big\{\rhobo\in (0,\infty )^2 \,\big|\,\Rbo (\rhobo )=\rhobo_c \big\}
\ee
denotes the set of all densities having background density $\rhobo_c \in\partial D_\rho$, and (\ref{subdiff}) implies that this is a linear set in normal direction to $D_\mu$. By Lemma \ref{theo21}, the normal vectors to $\partial D_\mu$ have two nonnegative components. So a first direct consequence of Theorem \ref{theo33} is that (as expected)
\be
\Rbo_c (\rhobo )\leq\rhobo\quad\mbox{i.e.}\quad R_{c,i} (\rhobo )\leq\rho_i \quad\mbox{for }i=1,2\ .
\ee
We say that species~$i$ condenses if $R_{c,i} (\rhobo )<\rho_i$\label{rhoc2} and define
\be\label{aidef}
A_i =\big\{\rhobo\in (0,\infty )^2 \,\big|\, R_{c,i} (\rhobo
)<\rho_i \big\}\subset (0,\infty )^2 \setminus D_\rho\ .
\ee
So the phase space $(0,\infty )^2$ can be partitioned in the following way:
\begin{itemize}
\item homogeneous phase region $D_\rho$, $\Rbo_c (\rhobo )=\rhobo$,
\item condensed phase region $A_1 \setminus A_2$, condensation of species 1 only,
\item condensed phase region $A_2 \setminus A_1$, condensation of species 2 only,
\item condensed phase region $A_1 \cap A_2$ condensation of both species,
\end{itemize}
defining the phase diagram of the model,
\be
PD=\big\{ D_\rho ,A_1 \setminus A_2 ,A_2 \setminus A_1 ,A_1 \cap A_2 \big\}\ .
\ee
The topology of the phase regions depends on the weight $w$ and all phases except $D_\rho$ may also be empty. Examples are given below in Figures \ref{fig1} and \ref{fig2} and in more detail also in Section 5.

The entropy density $s$ and the pressure $p$ are convex conjugates (\ref{entropy}), and thus $\rhobo\in\delta p(\mubo )\Leftrightarrow\mubo\in\delta s(\rhobo )$ (see Theorem \ref{duality}). Together with Theorem \ref{theo33} this implies that $s$ is an affine function on $\Rbo_c^{-1} (\rhobo_c )$ for every $\rhobo_c \in\partial D_\rho$, i.e.
\be
s(\rhobo )=s(\rhobo_c ) +(\rhobo -\rhobo_c )\Mbo (\rhobo_c )\quad\mbox{for all }\rhobo\in\Rbo_c^{-1} (\rhobo_c )\ .
\ee
So $s$ has a non-strictly supporting hyperplane in $\rhobo_c$, i.e.\ we have partial equivalence of ensembles in the sense of \cite{touchetteetal04}.

By $\overline\Mbo$ every condensed phase region is mapped on $\partial D_\mu \cap D_\mu$. Due to their definition, $A_1 \setminus A_2$ and $A_2 \setminus A_1$ correspond to special parts of that boundary. Define
\be\label{decom1}
\partial^i D_\mu :=\overline{\big\{\mubo\in\partial D_\mu \,\big|\,\partial D_\mu\mbox{ diff'able in }\mubo ,\, \nbo_\mubo \parallel \ebo_i \big\} }\quad\mbox{for }i=1,2\ ,
\ee
where $\ebo_i$ denotes the unit vector in direction $i$. By regularity of $D_\mu$ given in Lemma \ref{theo21}, namely convexity and completeness, $\partial^1 D_\mu$ (if non-empty) is a straight line of the form
\be\label{decom2}
\partial^1 D_\mu =\big\{\mubo^1 +\lambda\ebo_2 \,\big|\,\lambda\in\R\big\}\quad\mbox{or}\quad\partial^1 D_\mu =\big\{\mubo^1 -\lambda\ebo_2 \,\big|\,\lambda\geq 0\big\}\ ,
\ee
for some $\mubo^1 \in\partial D_\mu$. In the first case $\partial^2 D_\mu$ has to be empty, and an analogous version of (\ref{decom2}) holds for $\partial^2 D_\mu$. The connection with the phase region is basically a direct consequence of Theorem \ref{theo33}, summarized in the following.

\begin{corollary}\label{theo34}
The phase regions $D_\rho$, $A_1 \setminus A_2$ and $A_2 \setminus A_1$ are simply connected, $\overline\Mbo (A_i \setminus A_j )=\partial^i D_\mu \cap D_\mu$ and 
\be\label{decom3}
A_i \setminus A_j =\Rbo (\partial^i D_\mu \cap D_\mu )+\{\lambda\ebo_i |\lambda >0\}\quad\mbox{for }i=1,2\, ,\ j\neq i\ .
\ee
Here we use the convention that $M_1 +M_2 =\{ m_1 +m_2 |m_i \in M_i \}$ which is empty if one of the sets $M_i$ is empty.\\
$D_\rho \neq\emptyset$ and it is connected to $(0,0)$. If $A_i \setminus A_j \neq\emptyset$, it is connected to $\{\rhobo\, |\,\rho_j =0\}$.
\end{corollary}

Using this, the phase diagram is uniquely specified by $\partial D_\mu$ up to the exact location of $\partial D_\rho =\Rbo (\partial D_\mu \cap D_\mu )$. The phase region $A_1 \cap A_2$ corresponds to a more complicated part of the boundary. It may in general be disconnected and is most easily found as the complement of the union of all other phase regions. As a direct consequence of (\ref{decom3}), the phase boundary between $A_1 \cap A_2$ and $A_i \setminus A_j$, if both are non-empty, is a straight line in direction $\ebo_i$. These results are illustrated in Figure \ref{fig1} and \ref{fig2} for
\be\label{stawe}
w(\kbo )=\frac{k_1 !}{(1+b)_{k_1}}\,\Big(\frac{k_1 +1}{k_1 +2}\Big)^{k_2}\ ,
\ee
where $(1+b)_{k_1} =\prod_{i=0}^{k_1 -1} (i+b)$ denotes the Pochhammer symbol. This is the stationary weight of a zero-range process that has been introduced in \cite{evansetal03}, where also the phase diagrams are derived. This derivation and the corresponding zero-range process will be revisited in Section 5.
\bef
\begin{center}
\includegraphics[width=0.45\textwidth]{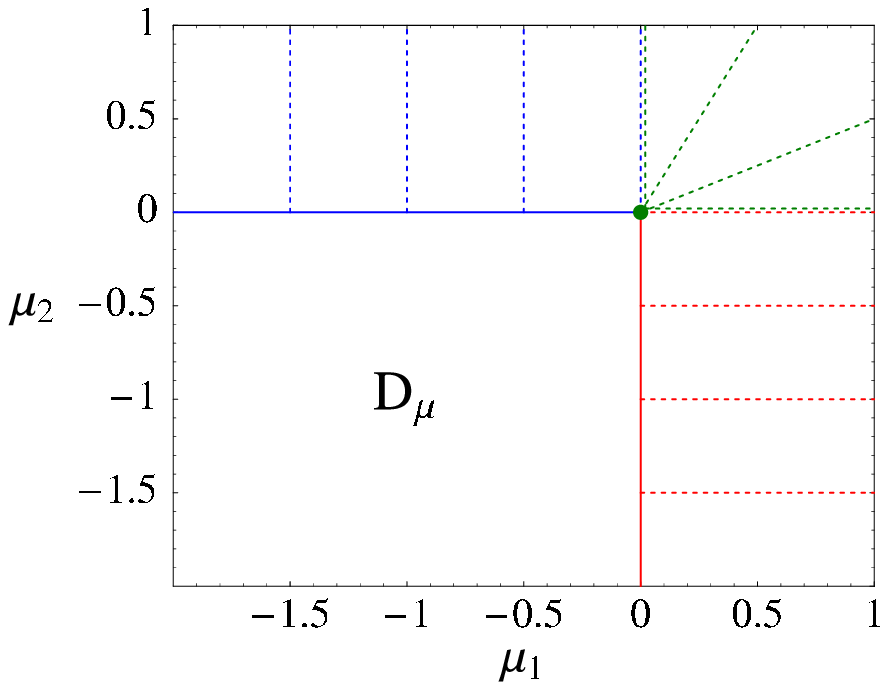}\hfill\includegraphics[width=0.45\textwidth]{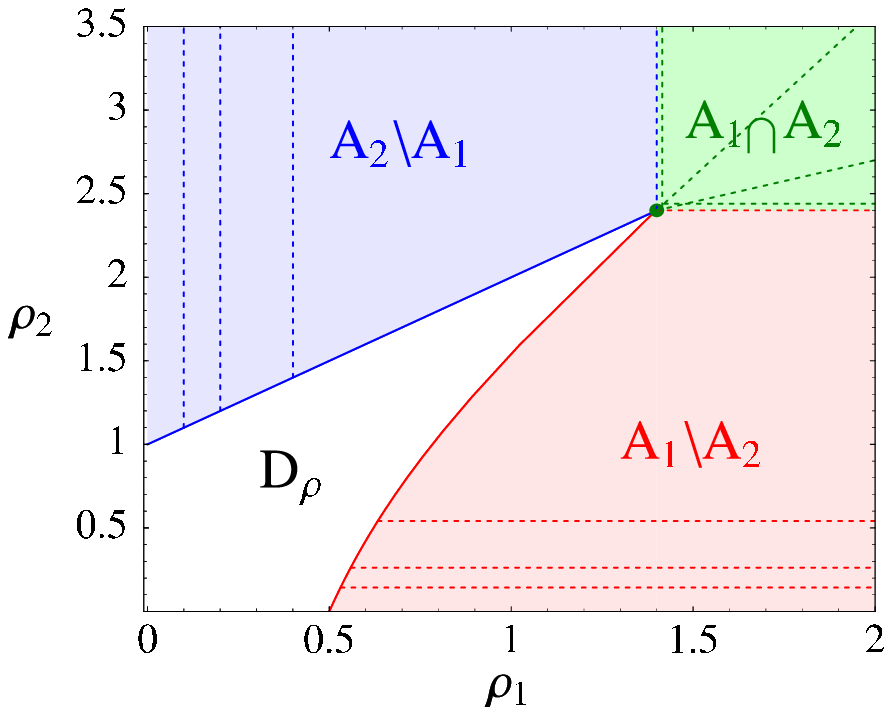}
\end{center}
\caption{$D_\mu$ and phase diagram for $w$ as given in (\ref{stawe}) with $b=4$. Dashed lines on the left denote the normal directions to $\partial D_\mu$ and on the right they determine the function $\Rbo_c (\rhobo )$ and the phase regions as given in Corollary \ref{theo34}.\label{fig1}}
\enf
\bef[t]
\begin{center}
\includegraphics[width=0.35\textwidth]{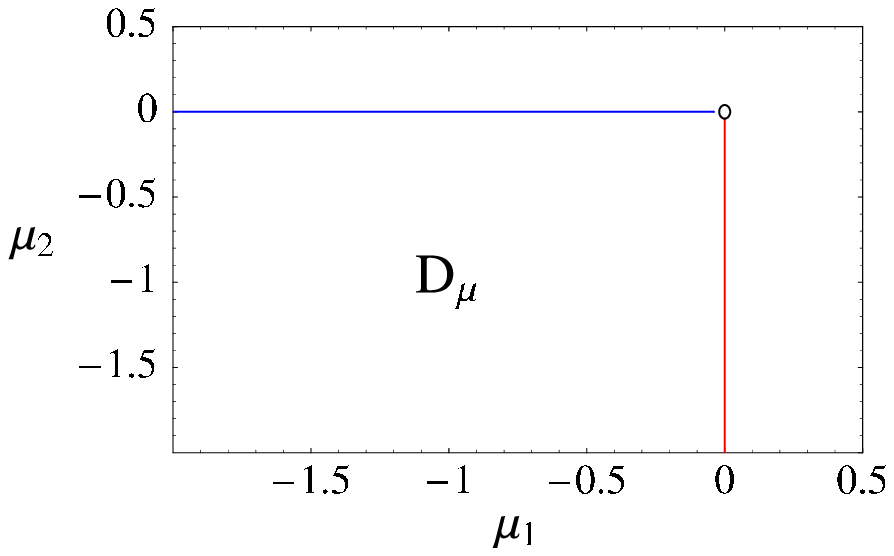}\hspace*{30mm}\includegraphics[width=0.35\textwidth]{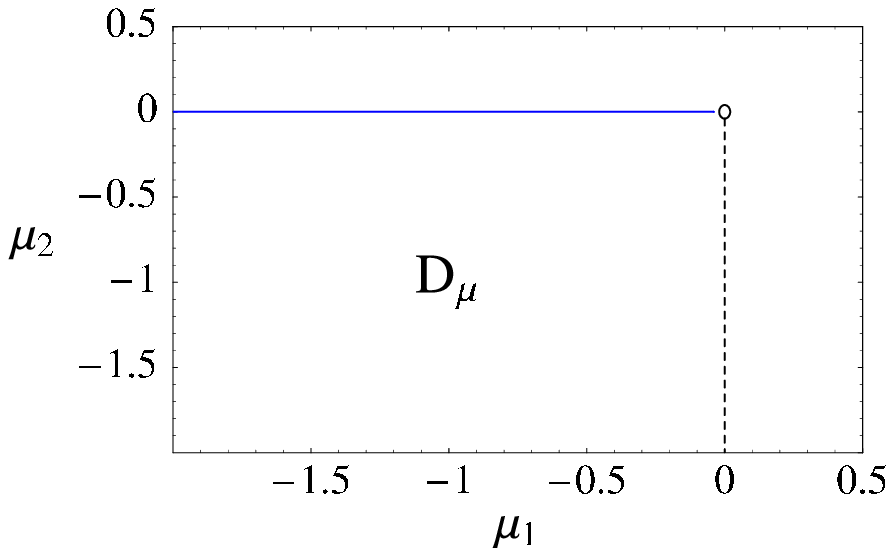}\\[2mm]
\hspace*{5mm} $b=3$\hspace*{66mm} $b=2$\\[5mm]
\includegraphics[width=0.45\textwidth]{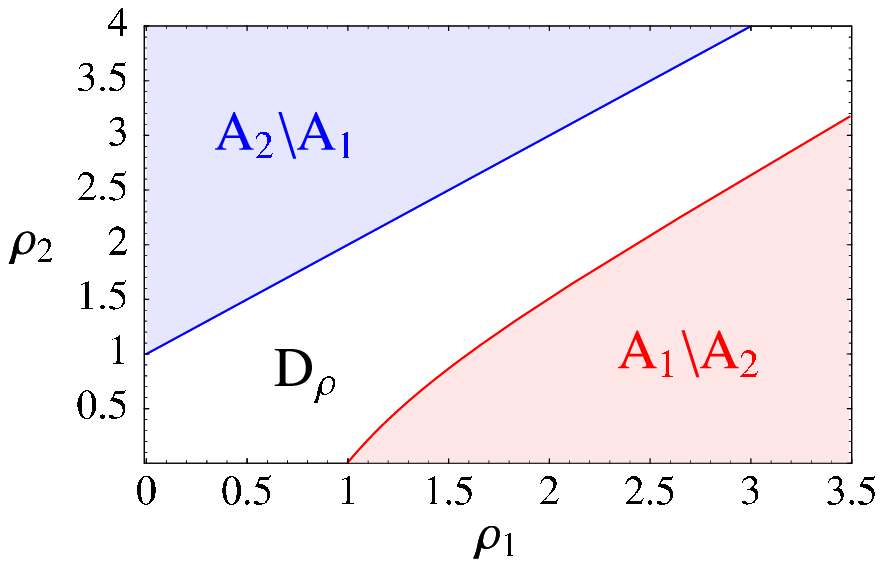}\hfill\includegraphics[width=0.45\textwidth]{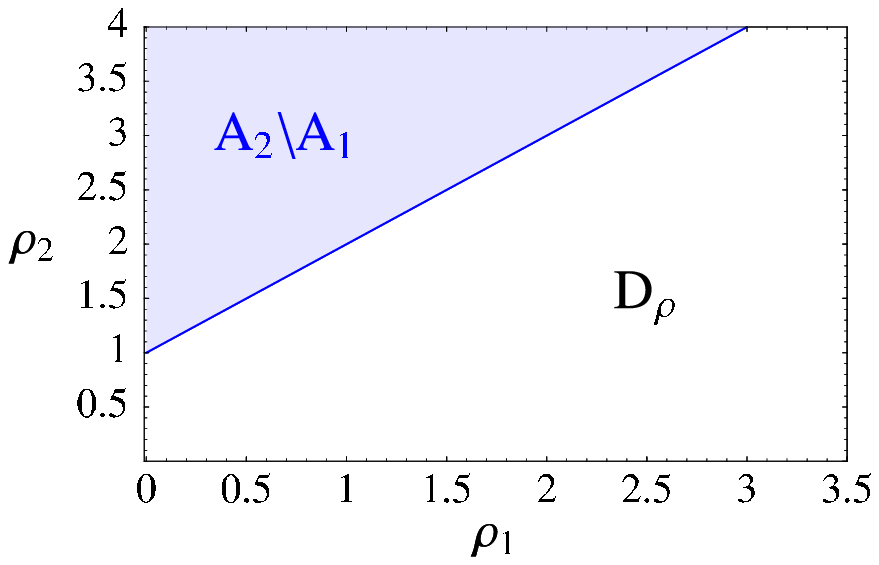}
\end{center}
\caption{$D_\mu$ and phase diagram for $w$ as given in (\ref{stawe}) with $b\leq 3$, which implies $A_1 \cap A_2 =\emptyset$ (for the relevant calculations see Section 5). For $b=3$ phase $A_1 \setminus A_2$ is non-empty as shown on the left, for $b=2$ only species 2 condenses as shown on the right.\label{fig2}}
\enf

By Proposition \ref{theo22}, occurrence of condensation can be characterized by boundary properties of $D_\mu$. For single species systems this can be directly translated to a condition on the weight $w$. $\partial D_\mu =\{\mu_c \}$ is only a single point and $\mu_c \in D_\mu$ if and only if
\be\label{single}
\nu^1_{\mu_c} (k)\propto w(k)\, e^{\mu_c k}=o(k^{-2})\ ,
\ee
and $\nu^1_{\mu_c}$ has finite first moment $\rho_c$, the critical density. For a two species system the condition $\partial D_\mu \cap D_\mu \neq\emptyset$ cannot be rephrased as a simple condition on $w$, and in general knowledge of $D_\mu$ is required. But for specific systems this is usually not difficult to obtain, and combining the results of this section, this is also sufficient to derive the phase diagram. This is explained in more detail for some generic examples in Section 5.

  \subsection{Properties of the condensed phase}
Note that the canonical measure can be written in conditional form
\be\label{condform}
\pi_{L,\Nbo_L} =\nu_\mubo^L \big(\, .\,\big|\,\{\Sigmabo_L =\Nbo_L\}\big)
\ee
for all $\mubo\in D_\mu$, and in particular for $\mubo =\overline\Mbo (\rhobo )$ where $\rhobo =\lim\limits_{L\to\infty} \Nbo_L /L$. For $\rhobo\not\in D_\rho$, $\{\Sigmabo_L =\Nbo_L \}$ is a large deviation event for $\nu_{\overline\Mbo (\rhobo )}^L$. A typical configuration for $\pi_{L,\Nbo_L }$ then corresponds to a most likely event to realize this large deviation.

Theorem \ref{theo31} implies that in such a configuration for large $L$ we have coexistence of a critical background domain with density $\Rbo_c (\rhobo )$ and a condensate containing $\big(\rhobo -\Rbo_c (\rhobo )\big) L$ particles on average. The number of particles in the condensate fluctuates, and due to the conservation law the distribution is given by the fluctuations of the number of particles in the background domain. The latter has extensive volume and the distribution converges to a product measure, so the particle number is approximately given by a sum of iidrv's. Thus if $\nu_{\overline\Mbo (\rhobo )}$ has finite second moment the central limit theorem applies and the fluctuations should be Gaussian. If the second moment is infinite the fluctuations are non-Gaussian and determined by the L\'evy stable law of $\nu_{\overline\Mbo (\rhobo )}$ \cite{rvaceva54}. This picture is consistent with results on single-species zero-range processes \cite{evansetal05b}. In general, the background distribution for each condensing species is subexponential, so non-Gaussian fluctuations are indeed possible.

\begin{lemma}\label{theo35}
Suppose $A_i \neq\emptyset$. The single site marginal $\nu_\mubo^{1,(i)} (k)=\sum\limits_{\kbo ,k_i =k} \nu_\mubo^1 (\kbo )$ with respect to species $i$ has a subexponential tail, i.e.
\be\label{state}
\liminf_{k \to\infty} -\frac{1}{k}\log\nu_\mubo^{1,(i)}(k )=0\ ,
\ee
if and only if $\mubo\in \overline{\Mbo}(A_i )\subset \partial D_\mu \cap D_\mu$. This holds with $\lim$ instead of $\liminf$ if $w$ has a regular tail as given in (\ref{cond2}). In this case there also exists a subexponential sequence in directions normal to $\partial D_\mu$. More precisely, for all $\nbo_\mubo$ normal\footnote{If $\partial D_\mu$ is differentiable in $\mubo$, $\nbo_\mubo$ is unique. Otherwise the statement holds for the two limiting normal directions $\nbo^+_\mubo$ and $\nbo^-_\mubo$ (cf. Theorem \ref{theo33}).} to $\partial D_\mu$ at $\mubo$ and for every sequence $\kbo_n$ with $|\kbo_n |\to\infty$ and $\kbo_n /|\kbo_n |\to\nbo_\mubo$ as $n\to\infty$ we have
\be\label{state2}
\lim_{n\to\infty} -\frac{1}{|\kbo_n |}\log\nu^1_\mubo \big(\kbo_n \big) =0\ .
\ee
\end{lemma}
The statement (\ref{state2}) is particularly important for the proof of Theorem \ref{theo31}. It is remarkable that a subexponential sequence exists precisely in the direction normal to $\partial D_\mu$. Together with (\ref{subdiff}) this allows all the excess mass in the system to condense on a single lattice site. If $\partial D_\mu$ is not differentiable in $\mubo$, (\ref{state2}) should also hold in all intermediate directions between $\nbo^+_\mubo$ and $\nbo^-_\mubo$. This is supported by the examples we studied so far (some of which are given in Section 5), which all show condensation on a single lattice site. However we are not able to prove this in general and it is also not required to prove Theorem \ref{theo31}. Statements similar to (\ref{state2}) on the limit behaviour of indices of multivariate generating functions have been derived only recently in the combinatorics literature based on Cauchy's integral formula (see \cite{pemantleetal05} and references therein).

Regarding the spatial extension of the condensate domain in large finite systems of size $L$, Theorem \ref{theo33} only assures that it covers a non-extensive volume of $o(L)$ sites. Suppose the tail of the single site marginal with respect to a condensing species is a power law with finite first moment, i.e.
\be\label{tail}
\nu_\mubo^{1,(i)} (k)\sim k^{-b}\quad\mbox{for some }b>2\mbox{ as }k\to\infty\ .
\ee
Then the following statement holds.

\begin{theorem}\label{theo36}
For $\rhobo\in A_i$, $\mubo =\overline\Mbo (\rhobo )$ and with (\ref{tail}) we have a weak law of large numbers, i.e. for all $\epsilon >0$
\be\label{lln}
\tilde\nu_\mubo^{L,(i)} \Big(\Big|\frac1L\,\max_{x\in\Lambda_L} \eta_i (x)- \big(\rho_i -R_{c,i} (\rhobo )\big)\Big| >\epsilon\Big)\to 0\quad\mbox{as }L\to\infty\ ,
\ee
where $\tilde\nu_\mubo^{L,(i)}=\nu_\mubo^{L,(i)} \big(\, .\,\big|\,\{\Sigma^i_L =N_{i,L}\}\big)$ with $N_{i,L} /L\to\rho_i$.
\end{theorem}
For single-species systems $\pi_{L,N_L} =\nu_\mu^L \big(\, .\,\big|\,\{\Sigma_L =N_L\}\big)$ and the theorem implies convergence with respect to the canonical distribution $\pi_{L,N_L}$ \cite{stefan}. Therefore in typical configurations for large $L$ all the excess mass concentrates on the site with maximal occupation number, so the condensate covers only a single lattice site. Since the particle system is translation invariant, the location of this site is chosen uniformly at random. This property is typical for large deviations of distributions with subexponential tails. The results in \cite{baltrunasetal02,vinogradov} and Monte-Carlo simulations of zero-range processes suggest that Theorem \ref{theo36} also holds for other subexponential tails. The proof given in Section 4.4, however, which is a slight extension of a result in \cite{jeonetal00}, works only for power law tails. In the case of two or more species, $\pi_{L,\Nbo_L }\neq\nu_\mubo^{L,(i)} \big(\, .\,\big|\,\{\Sigma^i_L =N_{i,L}\}\big)$. However, Monte-Carlo simulations of zero-range processes \cite{stefan3} confirm that the above implication of Theorem \ref{theo36}, which is rigorous for a single species, also holds for two-species systems.\\

\section{Proofs}
  \subsection{Proofs of Section 2}

\textbf{Proof of Lemma \ref{theo21}.} Note that $D_\mu =\dom\Rbo =\dom R_1 \cap\dom R_2$ and
\be\label{risum}
R_i (\mubo )=\frac{1}{z(\mubo )} \sum_{\kbo\in\N^2} k_i \, w(\kbo )\, e^{\mubo\cdot\kbo}\ .
\ee
By definition, $z(\mubo )>0$ so $\dom R_i$ is the domain of convergence of the sum in the numerator. Suppose $\dom R_i$ is non-empty and contains $\mubo^1 \neq\mubo^2$. For each $\mubo =\lambda\mubo^1 +(1-\lambda )\mubo^2$, $\lambda\in (0,1)$, by convexity of the exponential,
\be
e^{\mubo\cdot\kbo} \leq\lambda\, e^{\mubo^1\cdot\kbo}+(1-\lambda )\, e^{\mubo^2\cdot\kbo}
\ee
for all $\kbo\in\N^2$. So each term in the sum (\ref{risum}) can be bounded and we get
\be
\sum_{\kbo\in\N^2} k_i w(\kbo )e^{\mubo\cdot\kbo} \leq\lambda\sum_{\kbo\in\N^2} k_i w(\kbo )e^{\mubo^1\cdot\kbo} +(1{-}\lambda )\sum_{\kbo\in\N^2} k_i w(\kbo )e^{\mubo^2\cdot\kbo} <\infty\ ,
\ee
so $\mubo\in\dom R_i$. Thus $D_\mu$ is the intersection of two convex sets and is itself convex.
Since $\sum_{\kbo\in\N^2} k_i \, w(\kbo )\, e^{\mubo\cdot\kbo}$ is monotonically increasing in $\mu_1$ and $\mu_2$, $\mubo^* \in D_\mu$ implies $\mubo\in D_\mu$ for all $\mubo\in\Delta (\mubo^* )$ and $D_\mu$ is complete. Replacing $k_i w(\kbo )$ by $w(\kbo )$, both arguments also hold for $\dom z$, the domain of convergence of $\sum_{\kbo\in\N^2} w(\kbo )\, e^{\mubo\cdot\kbo}$. So $\dom z$ is convex and complete, and certainly $D_\mu \subset \dom z$.

To get a representation of the boundary we change variables and write
\be
R_i (\mubo )=\frac{1}{z(\mubo )}\sum_{m=0}^\infty \sum_{k_1 +k_2 =m} k_i \, w(\kbo )\, e^{\tilde\mu_1 (k_1 -k_2 )/2} e^{\tilde\mu_2 m/2} \ ,
\ee
with $\tilde\mu_1 =\mu_1 -\mu_2$ and $\tilde\mu_2 =\mu_1 +\mu_2$. The boundary of $D_\mu$ is then given by
\be
\tilde\mu_2 (\tilde\mu_1 )/2 =-\limsup_{m\to\infty} \frac1m\log\Big(\sum_{k_1 +k_2 =m} k_i \, w(\kbo )\, e^{\tilde\mu_1 (k_1 -k_2 )/2}\Big)\ .
\ee
Since for every positive function $f$
\be
\sup_{k_1 +k_2 =m}  f(\kbo )\leq\sum_{k_1 +k_2 =m} f(\kbo )\leq (m+1)\sup_{k_1 +k_2 =m}  f(\kbo )\ ,
\ee
and the logarithm is monotone we have
\be
\tilde\mu_2 (\tilde\mu_1 ) =-\limsup_{|\kbo |\to\infty} \Big( 2\log w(\kbo ) +\tilde\mu_1 (k_1 -k_2 )\Big)\big/ (k_1 +k_2 )\ .
\ee
This is independent of $i=1,2$ because $\log k_i /(k_1 +k_2)\to 0$, proving (\ref{ddmu}). In particular, replacing $k_i w(\kbo )$ by $w(\kbo )$ leads to the same formula for $\dom z$. So $D_\mu$ and $\dom z$ can only differ on the boundary, which implies ${\rm int}\,\dom z=\Di_\mu$.

Since $D_\mu$ is convex, $\tilde\mu_2$ is piecewise differentiable (see e.g. \cite{santalo}, Chapter 1). By completeness of $D_\mu$ it is clear that $D_\mu \neq\emptyset$ if and only if $\tilde\mu_2 (0)>-\infty$, which is equivalent to
\be\label{cond1b}
\limsup_{|\kbo |\to\infty}\frac{1}{k_1 +k_2 }\,\log w(\kbo )
=\inf_{n\in\N}\sup_{|\kbo |\geq n}\frac{1}{k_1 +k_2 }\,\log w(\kbo )<\infty\ .
\ee
This in turn is equivalent to (\ref{cond1}), because if (\ref{cond1}) holds (\ref{cond1b}) is bounded by $\xi <\infty$, and if (\ref{cond1}) does not hold there exists a sequence $\kbo_n$ for which $\frac{1}{|\kbo_n |}\,\log w(\kbo_n )$ is unbounded. The same argument works directly for $\dom z$.\hfill $\Box$\\

\textbf{Proof of Lemma \ref{theo21b}.} The existence of some finite exponential moments follows from the identity
\be
\langle e^{\thetabo\cdot\feta}\rangle_{\nu^1_\mubo} =Z(\mubo
+\thetabo )/Z(\mubo )
\ee
which is finite for sufficiently small $\thetabo\in\R^2$ if and only if $\mubo\in\Di_\mu$. Therefore $p\in C^\infty (\Di_\mu ,\R )$ and the covariance matrix
\be\label{cov1}
D^2 p(\mubo )=D\Rbo (\mubo )=\mathit{Cov} (\nu^1_\mubo )=\big(\langle\eta_i
\,\eta_j \rangle^c_{\nu^1_\mubo}\big)_{i,j=1,2} ,
\ee
is well defined on $\Di_\mu$, where $\langle\eta_i \,\eta_j \rangle^c_{\nu^1_\mubo} :=\langle\eta_i \,\eta_j \rangle_{\nu^1_\mubo}
-\langle\eta_i \rangle_{\nu^1_\mubo} \langle\eta_j
\rangle_{\nu^1_\mubo}$. Further, $D^2 p(\mubo )$ is symmetric
and positive definite, because
\be\label{convlogz}
\abo^T \cdot\big( D^2 p(\mubo )\big)\,\abo
=\big\langle (\abo\cdot (\eta_1 ,\eta_2 ))^2\big\rangle^c_{\nu^1_\mubo} >0
\ee
for all $\abo\in\R^2$ with $|\abo | =1$. Hence the eigenvalues of
$D^2 p(\mubo )$ are real and positive, which ensures that $p$
is strictly convex and $\Rbo$ is invertible on $\Di_\mu$.

Moreover, by completeness of $D_\mu$, one-sided derivatives $\partial_{\mu_i} p$ exist on $\partial D_\mu \cap D_\mu$, so $p\in C^1 (D_\mu ,\R )$. Strict convexity of $p$ extends to $\partial D_\mu \cap D_\mu$, since for any linear subset of the boundary $\mubo +\lambda\ebo_\phi$ in direction $\ebo_\phi$, parametrized by $\lambda$,
\be
\partial_\lambda p(\mubo +\lambda\ebo_\phi ) =\ebo_\phi\cdot\Rbo (\mubo +\lambda\ebo_\phi )\quad\mbox{ is strictly increasing with }\lambda\ .
\ee
This holds because by completeness of $D_\mu$, $\phi\in [\pi /2,\pi ]$ (or equivalently $[3\pi /2,2\pi ]$), and for $\phi\in (\pi /2,\pi ]$ and $\phi\in [\pi /2,\pi )$, $-R_1$ and $R_2$, respectively, are strictly increasing with $\lambda$ along $\mubo +\lambda\ebo_\phi$.

With a change of variables $\psi_i =e^{\mu_i } \in (0,\infty )$, $i=1,2$ we get
\be
z(\mubo )=\tilde z(\psibo )=\sum_{\kbo\in\N^2} w(\kbo )\,\psi_1^{k_1} \,\psi_2^{k_2}\ .
\ee
The domain of convergence of this two dimensional power series certainly contains points with $\psi_1 =0$ or $\psi_2 =0$, corresponding to $\mu_1 =-\infty$ or $\mu_2 =-\infty$, respectively. So the definition of $z$ can be extended to $\partial^{i,-\infty} D_\mu$, $i=1,2$, and thus also the definition of $p$, as $\tilde z(\psibo )\geq W(0,0)>0$ for all $\psibo$. Since $\tilde z$ is continuous on its domain of convergence and $0^{k_1} =\delta_{k_1 ,0}$,
\be
\lim_{\thetabo \to (-\infty ,\mu_2 )} z(\thetabo )=\tilde z(0,\psi_2 )=\sum_{k_2 =0}^\infty w(0,k_2 )\,\psi_2^{k_2}
\ee
which yields the result for $p(-\infty ,\mu_2 )$, and analogously $p(\mu_1 ,-\infty )$. An analogous argument holds for $\Rbo$, where
\be
R_2 (-\infty ,\mu_2 )=\frac{1}{z(-\infty ,\mu_2 )}\sum_{k_2 =0}^\infty k_2 \, w(0,k_2 )\, e^{\mu_2 k_2}
\ee
can be written as $R_2 (-\infty ,\mu_2 )=\partial_{\mu_2} p(-\infty ,\mu_2 )$, and $R_1 (-\infty ,\mu_2 )=0$ for all $\mu_2$, since $k_1 \, 0^{k_1} =0$ for all $k_1$.

By completeness of $D_\mu$, there exists $\mu_2^* \in (-\infty ,\infty]$, such that $\partial^{1,-\infty} D_\mu =\{ -\infty\}\times [-\infty ,\mu_2^* )$ or $\{ -\infty\}\times [-\infty ,\mu_2^* ]$, where the second case is only possible if $\mu_2^* <\infty$. Again by completeness, $D_\mu$ is bounded in $\mu_2$ if and only if $\mu_2^* <\infty$.\\
Now we can proceed analogous to \cite{kipnislandim}, Chapter 2, Lemma 3.3: Let $D_\mu$ be unbounded in $\mu_2$, i.e. $\partial^{1,-\infty} D_\mu =\{ -\infty\}\times [-\infty ,\infty )$. Suppose $R_2 (-\infty ,\mu_2 )=\partial_{\mu_2} p(-\infty ,\mu_2 )\leq C$ for all $\mu_2 \in [-\infty ,\infty )$. Then
\be
\log\frac{z(-\infty ,\mu_2 )}{z(-\infty ,0)}=p(-\infty ,\mu_2 )-p(-\infty ,0)\leq C\,\mu_2 \ ,
\ee
and thus $z(-\infty ,\mu_2 )\leq z(-\infty ,0)\, e^{C\mu_2}$ as $\mu_2 \to\infty$. This is in contradiction to $z$ being a power series with positive coefficients, and so $R_2$ is unbounded. Since $R_2$ is also monotone increasing on $\partial^{1,-\infty} D_\mu$ this implies $\lim\limits_{\mu_2 \to\infty } R_2 (-\infty ,\mu_2 )=\infty$, and $D_\rho$ is unbounded in $\rho_2$.\hfill $\Box$\\

\textbf{Proof of Proposition \ref{theo22}.} The proof uses basic results of convex analysis which are summarized in Appendix A. For fixed $\rhobo\in (0,\infty )^2$, the function $F:\R^2 \to (-\infty ,\infty ]$,
\be\label{fdef}
F(\mubo )=\left\{\bacl p(\mubo )-\rhobo\cdot\mubo\ &,\ \mubo\in\dom z\\ \infty\ &,\ \mubo\not\in\dom z \ea\right.\ \ ,
\ee
is strictly convex on $D_\mu \subset\dom z$. So by Theorem \ref{conmin}, ${\rm argmin}\, F\cap D_\mu$, the set of minimizers on $D_\mu$ contains at most one element. By (\ref{entropy}), $s(\rhobo )=-\inf_{\mubo\in D_\mu} F(\mubo )$, so $\overline\Mbo (\rhobo )$ is a minimizer of $F$. Also $F\in C^1 (D_\mu ,\R )$ (by regularity of $p$), and
\be\label{fgrad}
\grad F(\mubo )=\rhobo -\Rbo (\mubo )\ .
\ee
Thus for $\rhobo\in D_\rho$ we have $\grad F\big(\Mbo (\rhobo )\big) =0$ and $F$ has a local minimum in $\Mbo (\rhobo )$. By Theorem \ref{conmin} this is then the unique global minimum of $F$ on $D_\mu$. On the other hand, if $\rhobo\not\in D_\rho$, $\grad F(\mubo )\neq (0,0)$ for all $\mubo\in D_\mu$, and if there exists a minimizer for $F$ it has to be in $\partial\dom z$. In the following we will show that this minimizer exists and lies in fact in $\partial D_\mu \cap D_\mu$.

First we show that
\be\label{firstshow}
\inf_{\mubo\in\dom z} F(\mubo )=\inf_{\mubo\in E} F(\mubo )\quad\mbox{for some compact, non-empty }E\subset D_\mu \ .
\ee
Since $F$ is convex and $C^1$ we have for a fixed $\mubo^* \in \Di_\mu$
\be
F(\mubo )\geq F(\mubo^* )+\grad F(\mubo^* )\cdot (\mubo -\mubo^* )=F(\mubo^* )+\big(\Rbo (\mubo^* )-\rhobo\big)\cdot (\mubo -\mubo^* )\ ,
\ee
for all $\mubo\in\R^2$, following (\ref{subgr}). For any $\rhobo\in (0,\infty )^2$, there exists $\mubo^* \in\Di_\mu$ such that $R_i (\mubo^* )<\rho_i$, $i=1,2$, since $\Rbo (\mubo )\to (0,0)$ as $\mubo\to (-\infty ,-\infty )$ and $D_\mu$ is complete and nonempty. Thus
\be\label{edef}
F(\mubo )\geq F(\mubo^* )\quad\mbox{for all }\mubo\in E^* :=\big\{\mubo :\big(\Rbo (\mubo^* )-\rhobo\big)\cdot (\mubo -\mubo^* )\geq 0\big\}\ .
\ee
Since both components of $\Rbo (\mubo^* )-\rhobo$ are negative, the half-space $E^*$ contains the cone $\Delta (\mubo^* )$, as given in (\ref{deltadef}). The boundary $\partial E^*$ is a straight line. Since $\mubo^* \in {\rm int}\,\dom z$ and $\dom z$ is convex, $\partial E^* \cap\partial\dom z$ consists of at most two points.
\begin{enumerate}
\item Suppose $\partial E^* \cap\partial\dom z$ has two points, then $\dom z\setminus E^*$ is bounded and contains the level set $E:=\lev_{\leq F(\mubo^* )} F$. By definition (\ref{fdef}), $F$ is lower semicontinuous and thus $E$ is closed by Theorem \ref{levels} and therefore compact, which implies (\ref{firstshow}).
\item $\partial E^* \cap\partial\dom z =\emptyset$ is only possible if $\partial\dom z \parallel\partial E^*$ is itself a line. By strict convexity of $F$ the level line $\lev_{=F(\mubo^* )} F$ touches $\partial E^*$ only in the single point $\mubo^*$. So we can find a point $\mu^{**}$ close to $\mu^*$ such that $\partial E^{**} \cap\partial\dom z$ contains one point, where $E^{**}$ is defined as in (\ref{edef}). Then continue as in case (3) below.
\item If $\partial E^* \cap\partial\dom z$ consists of one point, $\partial E^* \cap\dom z$ is a semi-infinite line and let $\ebo_\phi$ be the direction of that line in which there is no intersection point. Then $\phi\in (\pi /2 ,\pi)$ or $\phi\in (3\pi /2 ,2\pi)$, and in the first case $\dom z$ is unbounded in $\mu_2$ and in the second case it is unbounded in $\mu_1$. Concentrating on the second case, by Lemma \ref{theo21b} there exists $\mubo\in D_\mu$ such that $R_1 (\mubo )>\rho_1$, and thus by continuity of $F$ and (\ref{fgrad}) there has to be a point $\mubo^1 \in D_\mu$ such that $\partial_{\mu_1} F(\mubo^1 )>0$ and $\partial_{\mu_2} F(\mubo^1 )<0$. Then $\dom z\setminus (E^* \cup E^1 )$ is bounded, where $E^1$ is defined analogously to $E^*$ in (\ref{edef}). If $\mubo^1 \in E^*$ then $E:=\lev_{\leq F(\mubo^* )} F\subset \dom z\setminus (E^* \cup E^1 )$ is bounded and compact as in the first case (1). If $\mubo^1 \not\in E^*$ we take $E:=\lev_{\leq F(\mubo^1 )} F\subset \dom z\setminus (E^* \cup E^1 )$, which is also compact. The same argument works for $\phi\in (\pi /2 ,\pi)$, with $\mubo^2$ such that $\partial_{\mu_1} F(\mubo^2 )<0$ and $\partial_{\mu_2} F(\mubo^2 )>0$.
\end{enumerate}
This completes the proof of (\ref{firstshow}).

Since $F$ is continuous on $\dom z$, (\ref{firstshow}) implies that it has at least one minimizer $\bar\mubo$ on $\dom z$. By completeness of $\dom z$ and $D_\mu$, $\bar\mubo +t\ebo\in D_\mu$ for all $t<0$, where $\ebo =(1,1)/\sqrt{2}$. By strict convexity of $p$,
\be
\frac{d}{dt}p(\bar\mubo +t\ebo )=\Rbo (\bar\mubo +t\ebo )\cdot\ebo
\ee
is monotone increasing with $t$. Therefore $\lim_{t\nearrow 0} \Rbo (\bar\mubo +t\ebo )\cdot\ebo$ exists and is given by
\be
\frac{1}{z(\bar\mubo )}\lim_{t\nearrow 0} \sum_{\kbo\in\N^2} (k_1 +k_2 )\, e^{(\bar\mubo +t\ebo )\cdot\kbo} =R_1 (\bar\mubo )+R_2 (\bar\mubo )
\ee
by monotone convergence. Now suppose $\bar\mubo\not\in D_\mu$, then this implies
\be
\frac{d}{dt}F(\bar\mubo +t\ebo )=\big(\Rbo (\bar\mubo +t\ebo )-\rhobo\big) \cdot\ebo\to \infty\quad\mbox{as }t\nearrow 0\ .
\ee
This is a contradiction to $\bar\mubo$ being a minimizer of $F$, so $\bar\mubo\in D_\mu$. By the first part of the proof we have uniqueness on $D_\mu$, so $\overline\Mbo (\rhobo ):=\bar\mubo\in\partial D_\mu \cap D_\mu$ is the unique minimizer of $F$ for all $\rhobo\not\in D_\rho$. 

Since $F\in C^1 (D_\mu ,\R)$ is linear in $\rhobo$ and strictly convex on $D_\mu$ for all $\rhobo\in (0,\infty )^2$, its maximizer $\overline\Mbo (\rhobo )$ is a continuous function of $\rhobo$. Since $\Rbo$ and its inverse are continuous, $D_\mu$ and $D_\rho$ are diffeomorphic and $\partial D_\rho \cap D_\rho =\Rbo (\partial D_\mu \cap D_\mu )$. Suppose there exists $\rhobo\in\partial D_\rho \setminus D_\rho$. Then there exists $\rhobo_j \in D_\rho$ ,$j\in\N$, such that $\rhobo_j \to\rhobo$ as $j\to\infty$. But then by continuity of $\overline\Mbo$ and $\Rbo$
\be
\rhobo_j =\Rbo\big(\Mbo (\rhobo_j )\big) \to\Rbo\big(\overline\Mbo (\rhobo )\big)\in D_\rho\ ,
\ee
in contradiction to $\rhobo\not\in D_\rho$. Thus $D_\rho$ is closed in $(0,\infty )^2$ and $\partial D_\rho =\Rbo (\partial D_\mu \cap D_\mu )$.\hfill $\Box$\\

  \subsection{Proofs of Section 3.2}

\textbf{Proof of Theorem \ref{theo33}.} In the following suppose that $\partial D_\mu$ is differentiable in $\mubo_c$. In case it is not, analogous arguments hold with $\lambda\nbo$ replaced by $\lambda^+ \nbo^+ +\lambda^- \nbo^-$.

For every $\rhobo_c \in\partial D_\rho$ also $\rhobo_c \in D_\rho$ since $D_\rho$ is closed, and $\mubo_c =\Mbo (\rhobo_c )\in\partial D_\mu \cap D_\mu$ by Proposition \ref{theo22}. One-sided derivatives of $p$ exist in $\mubo_c$ and
\be
\grad p(\mubo_c )=\rhobo_c \in\delta p(\mubo_c )\ ,
\ee
where the definition of the subgradient is given in (\ref{subgr}). Let $\nbo\perp\partial D_\mu$ in $\mubo_c$. Then $\lambda\nbo\cdot (\mubo -\mubo_c )\leq 0$ for all $\lambda\geq 0$ and $\mubo\in D_\mu$, and thus
\be
p(\mubo )\geq p(\mubo_c )+(\rhobo_c +\lambda\nbo )\cdot (\mubo -\mubo_c )\ ,
\ee
i.e. (\ref{subgr}) holds and $\rhobo_c +\lambda\nbo\in\delta p(\mubo_c )$. On the other hand, if $\nbo\not\perp\partial D_\mu$ or $\lambda <0$ there exists $\mubo^* \in D_\mu$ with $\lambda\nbo\cdot (\mubo^* -\mubo_c )>0$ and $p(\mubo^* )\geq p(\mubo_c )$ (by strict convexity of $p$) and (\ref{subgr}) does not hold, so that $\delta p(\mubo_c )=\big\{\rhobo_c +\lambda\nbo\,\big|\,\lambda\geq 0\big\}$.

$\mubo_c =\Mbo (\rhobo_c )$ minimizes $p(\mubo )-\rhobo_c \cdot\mubo$ and thus minimizes also \be
p(\mubo )-\rhobo_c \cdot\mubo -\lambda\nbo\cdot (\mubo -\mubo_c )\geq p(\mubo_c )-\rhobo_c \cdot\mubo_c
\ee
for all $\lambda\geq 0$, with equality if and only if $\mubo =\mubo_c$. So $\overline\Mbo\big(\delta p(\mubo_c )\big) =\{\mubo_c \}$, $\Rbo_c \big(\delta p(\mubo_c )\big) =\{\rhobo_c \}$ and thus $\delta p(\mubo_c )\subset \Rbo_c^{-1} (\rhobo_c )$. On the other hand, $\rhobo\in\Rbo_c^{-1} (\rhobo_c )=\overline\Mbo^{-1} \big(\mubo_c )$ implies $\overline\Mbo (\rhobo )=\mubo_c$, and thus $s(\rhobo )=\rhobo\cdot\mubo_c -p(\mubo_c )$ by Proposition \ref{theo22}, which is equivalent to $\rhobo\in\delta p(\mubo_c )$ by Theorem \ref{duality}. Therefore $\delta p(\mubo_c )= \Rbo_c^{-1} (\rhobo_c )$, finishing the proof.\hfill $\Box$\\
\\
\textbf{Proof of Corollary \ref{theo34}.} Since $\big\{\mubo |\mu_i <\tilde\mu_2 (0)/2 ,\, i=1,2\big\}\subset D_\mu$ by the proof of Lemma \ref{theo21} and $\Rbo (\mu ,\mu )\to (0,0)$ as $\mu\to -\infty$, $D_\rho \neq\emptyset$ is connected to $(0,0)$. $D_\rho$ is simply connected since it is diffeomorphic to the convex set $D_\mu$. 

By definition of $A_i \setminus A_j$ in (\ref{aidef}) and Theorem \ref{theo33}, for each $\rhobo\in A_i \setminus A_j$, $\ebo_i$ is a normal vector to $\partial D_\mu$ at $\overline\Mbo (\rhobo )$, and thus $\overline\Mbo (\rhobo )\in\partial^i D_\mu \cap D_\mu$ by definition (\ref{decom1}). On the other hand, for every $\mubo\in\partial^i D_\mu \cap D_\mu$, $\overline\Mbo^{-1} (\mubo )\supset\Rbo (\mubo )+\{\lambda \ebo_i |\lambda >0\}$ by Theorem \ref{theo33} and thus we have shown (\ref{decom3}). Since $\partial^i D_\mu$ is either empty or a (simply connected) straight line according to (\ref{decom2}), the same holds for $\partial^i D_\mu \cap D_\mu$ by completeness of $D_\mu$, and thus also $A_i \setminus A_j$ is simply connected. If $\partial^i D_\mu \cap D_\mu$ is non-empty it is connected to $\mu_j =-\infty$ by (\ref{decom2}) for $j\neq i$, and thus $\Rbo (\partial^i D_\mu \cap D_\mu )$ is connected to $\{\rhobo |\rho_j =0\}$ since $R_j (\mubo )\to 0$ as $\mu_j \to -\infty$. Therefore by (\ref{decom3}) shown above, $A_i \setminus A_j$ is connected to $\{\rhobo |\rho_j =0\}$.\hfill $\Box$\\

  \subsection{Proof the equivalence of ensembles}

In the proof of Theorem \ref{theo31} we make use of Lemma \ref{theo35}, which is proved first.

\textbf{Proof of Lemma \ref{theo35}.}
If $\displaystyle\alpha =\liminf_{k\to\infty} -\frac{1}{k}\log\nu_\mubo^{1,(i)}(k)>0$ then for almost every $k$ it is $\nu_\mubo^{1,(i)} (k)\leq e^{-\alpha k}$. Thus
\be
\big\langle e^{\eta_i \alpha /2}\big\rangle_{\nu^{1,(i)}_\mubo} =\frac{z\big(\mubo +\ebo_i \alpha /2\big)}{z(\mubo )} <\infty\ ,
\ee
which is in contradiction to $\mubo\in \overline{\Mbo}(A_i )$ according to Lemma \ref{theo21}.

To prove the second statement of the lemma we use the integral criterion that for all $\mubo\in\R^2$, $z(\mubo )<\infty$ if and only if
\be\label{intcrit}
\tilde z(\mubo )=\int_{[0,\infty )^2}\!\!\! w(\xbo )\, e^{\mubo\cdot\xbo} \, d^2 x=\int_0^\infty\!\! \int_0^{\pi /2}\!\!\! w(r\ebo_\phi )\, e^{r\mubo\cdot\ebo_\phi}\, r\, d\phi\, dr<\infty\ .
\ee
By regularity of $w$ we can write the integral in polar coordinates $(r,\phi )$ with $r=\|\xbo\|_2$ and $\ebo_\phi =\xbo /r$. Depending on $w$, (\ref{intcrit}) is not necessarily equivalent to $z(\mubo )<\infty$. In this case, however, $w$ can be replaced by another function of the same regularity having the same values $w(\kbo )$, $\kbo\in\N^2$, for which equivalence holds. For example one could take $w$ to be a smoothed version of a piecewise linear interpolation of the points $w(\kbo )$, $\kbo\in\N^2$.

First assume that $\partial D_\mu$ is differentiable at $\mubo\in\partial D_\mu$ and pick a direction $\phi\in [0,\pi/2]$. Define the positive half space
\be
Q(\mubo ,\phi )=\big\{ \thetabo\,\big|\, (\thetabo -\mubo )\cdot\ebo_\phi >0\big\}\subset\R^2 \ .
\ee
Since $D_\mu$ is convex and $\partial D_\mu$ is differentiable in $\mubo$ we have
\be
Q(\mubo ,\phi )\cap\Di_\mu =\emptyset\quad\mbox{if and only if}\quad\ebo_\phi =\nbo_\mubo \perp\partial D_\mu \ .
\ee
Let $\psi\in [0,\pi /2]$ be the direction of $\nbo_\mubo$. Then for all $\phi\neq\psi$ there exists $\mubo_\phi \in Q(\mubo ,\phi )\cap\Di_\mu$, and
\be
w(r\ebo_\phi )\, e^{r\mubo\cdot\ebo_\phi} =w(r\ebo_\phi )\, e^{r\mubo_\phi \cdot\ebo_\phi}\, e^{r(\mubo -\mubo_\phi )\cdot\ebo_\phi}\ .
\ee
Since $\nu^1_{\mubo_\phi}$ has some exponential moments and $(\mubo -\mubo_\phi )\cdot\ebo_\phi <0$, there exists $\epsilon_\phi >0$ such that
\be
w(r\ebo_\phi )\, e^{r\mubo\cdot\ebo_\phi} =O(e^{-\epsilon_\phi r})\ .
\ee
So the integrand of (\ref{intcrit}) decays exponentially fast in all directions $\phi\neq\psi$. Further, for all $\phi\in [0,\pi /2]$ the maximal $\epsilon_\phi$ to choose is given by
\be
\tilde\epsilon_\phi =-\lim_{r\to\infty}\frac1r \log w(r\ebo_\phi ) -\ebo_\phi \cdot\mubo\ ,
\ee
The limit exists due to regularity of $w$ and further $\tilde\epsilon_\phi$ is a continuous function of $\phi$. We know that $\tilde\epsilon_\phi >0$ for all $\phi\neq\psi$ and thus in direction $\psi$ the integrand of (\ref{intcrit}) can decay only subexponentially, i.e. $\tilde\epsilon_\psi =0$. Otherwise there would exist $\thetabo$ with $\theta_i >0$ such that $\thetabo\cdot\ebo_\phi <\tilde\epsilon_\phi /2$ for all $\phi\in [0,\pi /2]$. By regularity of the integrand we could use Fubini's theorem to get
\be
\tilde z(\mubo +\thetabo )\leq\int_0^{\pi /2} \frac{1}{\tilde\epsilon_\phi -\thetabo\cdot\ebo_\phi}\, d\phi <\infty\ ,
\ee
which is in contradiction to $\mubo\in\partial D_\mu$. Alternatively, we also have $\tilde\epsilon_\phi \to 0$ as $\phi\to\psi$, since
\be\label{psilim}
\sup_{\thetabo\in Q(\mubo ,\phi )\cap D_\mu}|\thetabo -\mubo |\to 0\quad\mbox{as }\phi\to\psi\ ,
\ee
and $\tilde\epsilon_\psi =0$ follows by continuity of $\tilde\epsilon_\phi$.

This second argument also works if $\partial D_\mu$ is not differentiable in $\mubo$ for the two limiting normal directions $\psi^- <\psi^+$. (\ref{psilim}) holds as $\phi\nearrow\psi^-$ and $\phi\searrow\psi^+$, leading to $\tilde\epsilon_{\psi^-} =0$ and $\tilde\epsilon_{\psi^+} =0$, respectively. With continuity of $w$ the statement for $\nu_\mubo^1 (\kbo_n )$ follows for every sequence $\kbo_n$ with $|\kbo_n |\to\infty$ and $\kbo_n /|\kbo_n |\to\nbo_\mubo$.\hfill $\Box$\\

\textbf{Proof of Theorem \ref{theo31}.}
According to (\ref{forms}) the specific relative entropy is
\be
h_{L,\Nbo_L } \big(\overline\Mbo (\rhobo )\big) =-\frac1L\,\log\nu_{\overline\Mbo (\rhobo )}^L \big(\{\Sigmabo_L =\Nbo_L \}\big)\ ,
\ee
and we have to find a subexponential lower bound to the probability on the right-hand side.

For $\rhobo\in \Di_\rho$ we have $\overline{\Mbo} (\rhobo ) =\Mbo (\rhobo )\in D_\mu$ and $\nu^1_{\Mbo (\rhobo )}$ has exponential moments and thus finite covariance. The limit distribution of $\big(\Sigmabo_L (\feta )-\rhobo L\big) /\sqrt{L}$ is a bivariate Gaussian and we have $\nu_{\overline{\Mbo}(\rhobo )}^L \big(\{ \Sigmabo_L (\feta )=\Nbo_L \}\big)\sim 1/L$ as $L\to\infty$ by the multivariate local limit theorem, since $\Nbo_L /L\to\rhobo$. For $\rhobo\in\partial D_\rho$ the same argument holds as long as $\nu^1_{\Mbo (\rhobo )}$ has finite covariance. In case it does not, the same conclusion follows from convergence to an $\alpha$-stable limit law with rate of convergence $1/L^{2/\alpha} f(L)$, where $\alpha\in (0,2)$ and $f$ is slowly varying as $L\to\infty$ (see \cite{rvaceva54,mukhin91} for multivariate local limit theorems in the non-Gaussian case).

For $\rhobo\not\in D_\rho$ a typical number of particles under $\nu_{\overline\Mbo (\rhobo )}^L$ is $\big[\Rbo_c (\rhobo ) L\big]$, where $[..]$ denotes the integer value. We get a lower bound for $\nu_{\overline\Mbo (\rhobo )}^L \big(\{\Sigmabo_L =\Nbo_L \}\big)$ by a special configuration where the excess particles are concentrated on the first lattice site, i.e. $\feta (1)=\Nbo_L -\big[\Rbo_c (\rhobo ) L\big]$. Thus we have for all $\mubo\in D_\mu$
\be
\nu_\mubo^L \big(\{\Sigmabo_L =\Nbo_L \}\big)\geq\nu_\mubo^L \Big(\Big\{\Sigmabo_L =\Nbo_L ,\feta (1)=\Nbo_L -\big[\Rbo_c (\rhobo ) L\big]\Big\}\Big)\ .
\ee
First we assume that $\partial D_\mu$ is differentiable in $\overline\Mbo (\rhobo )$. By Theorem \ref{theo33}, $\rhobo -\Rbo_c (\rhobo )$ is perpendicular to $\partial D_\mu$ in $\overline\Mbo (\rhobo )$. Thus by Lemma \ref{theo35}
\bea\label{esti}
h_{L,\Nbo_L } \big(\overline{\Mbo}(\rhobo )\big)&\leq &-\frac1{L} \log \nu^1_{\overline{\Mbo}(\rhobo )} \Big( \Nbo_L -\big[\Rbo_c (\rhobo ) L\big]\Big)\nonumber\\
& &-\frac1L \log \nu_{\overline{\Mbo}(\rhobo )}^{L-1} \Big(\Big\{\Sigmabo_{L-1} =\big[\Rbo_c (\rhobo ) L\big]\Big\}\Big)\ \to\ 0
\eea
as $L\to\infty$, since $\big(\Nbo_L -[\Rbo_c (\rhobo ) L]\big) /L\to\big(\rhobo -\Rbo_c (\rhobo )\big)$. The second term vanishes with a local limit theorem analogous to the case $\rhobo\in\partial D_\rho$. If $\partial D_\mu$ is not differentiable in $\overline\Mbo (\rhobo )$ we may have two limiting normal vectors $\nbo^+_{\overline\Mbo (\rhobo )}$ and $\nbo^-_{\overline\Mbo (\rhobo )}$ which are linearly independent. Therefore there exist $\alpha^+_L ,\alpha^-_L \geq 0$ such that
\be\label{2sites}
\Nbo_L -\big[\Rbo_c (\rhobo ) L\big] =\big[\alpha^+_L \nbo^+_{\overline\Mbo (\rhobo )}\big] +\big[\alpha^-_L \nbo^-_{\overline\Mbo (\rhobo )}\big]
\ee
for all $L$. Then we proceed analogous to (\ref{esti}) and get a lower bound by distributing the excess particles on the first two lattice sites according to $\feta (1)=\big[\alpha^+_L \nbo^+_{\overline\Mbo (\rhobo )}\big]$ and $\feta (2)=\big[\alpha^-_L \nbo^-_{\overline\Mbo (\rhobo )}\big]$. By Lemma \ref{theo35} we have as $L\to\infty$,
\bea\label{estii}
h_{L,\Nbo_L} \big(\overline{\Mbo}(\rhobo )\big)&\leq &-\frac1{L} \log \nu^1_{\overline{\Mbo}(\rhobo )} \Big( \big[\alpha^+_L \nbo^+_{\overline\Mbo (\rhobo )}\big]\Big) -\frac1{L} \log \nu^1_{\overline{\Mbo}(\rhobo )} \Big( \big[\alpha^-_L \nbo^-_{\overline\Mbo (\rhobo )}\big]\Big)\nonumber\\
& &-\frac1L \log \nu_{\overline{\Mbo}(\rhobo )}^{L-2} \Big(\Big\{\Sigmabo_{L-2} =\big[\Rbo_c (\rhobo ) L\big]\Big\}\Big)\ \to\ 0\ .
\eea
\hfill $\Box$

We note that condition (\ref{cond2}) is not necessary to conclude (\ref{esti}) and (\ref{estii}). It would be enough to have some sequence $\kbo_n$ in direction $\rhobo -\Rbo_c (\rhobo )$ such that (\ref{state2}) holds in the limit $n\to\infty$ and $\sup_n \big|\kbo_{n+1} -\kbo_n \big| \leq C$ for some $C\in\R$. However this is still not a necessary condition and therefore we used the comparatively simple assumption (\ref{cond2}) which is fulfilled by a large number of examples.\\

\textbf{Proof of Corollary \ref{theo32}.}
The first statement (\ref{corr1}) with $f\in C(X,\R )$ such that $\langle e^{\epsilon f}\rangle_{\nu_{\overline\Mbo (\rhobo )}}<\infty$ for some $\epsilon >0$ is shown in \cite{csiszar75}, Lemma 3.1. It follows directly from subadditivity of relative entropy (see e.g. \cite{csiszar84}) and the inequality $ab<a\log a +e^b$, $a,b>0$, which leads to the variational formula \cite{varadhan88}
\be
H\big(\pi^n_{L,\Nbo} \big|\nu_\mubo^n \big) =\sup_{f\in C(X,\R )} \Big\{\langle f\rangle_{\pi^n_{L,\Nbo}} -\log\langle e^f \rangle_{\nu_\mubo^n}\Big\}
\ee
for the $n$-site marginals, $L\geq n$, $\Nbo\in\N^2$. Note that along with the product measure $\nu_\mubo$ also the canonical measures $\pi_{L,\Nbo}$ are permutation invariant, so that we do not have to specify the lattice sites in the $n$-site marginals. Following (\ref{forms}),
\be
\lim_{L\to\infty} \frac1L\,\log Z_{L,\Nbo_L}=p(\mubo )-\mubo\cdot\rhobo-\lim_{L\to\infty} h_{L,\Nbo_L} (\mubo )
\ee
for all $\mubo\in D_\mu$. Inserting $\mubo =\overline\Mbo (\rhobo )$ leads to (\ref{corr2}).\hfill $\Box$\\

  \subsection{Remaining proofs of Section 3.3}

The proof of Theorem \ref{theo36} uses large deviation results on the asymptotic distribution of $\Sigma_L^{(i)} (\feta )$ as $L\to\infty$, which we summarize in the following lemma for our purpose.

\begin{lemma}
Let $\omega_1 ,\omega_2 ,\ldots \in\ganz$ be i.i.d.\ random variables with mean $\rho_c$ and probability mass function $\nu (k)\sim k^{-b}$ with $b>2$. Then for every $\rho >\rho_c$ and some constant $c$, as $L\to\infty$,
\bea\label{ldresult}
\nu^L \Bigg(\bigg\{\sum_{x=1}^L \omega_x \geq \rho L\bigg\}\Bigg) &\simeq &L\,\nu\Big(\big\{\omega_1 \geq (\rho - \rho_c )L\big\}\Big)\ ,\\
\label{ldresult2}\nu^L \Bigg(\bigg\{\sum_{x=1}^L \omega_x = [\rho L]\bigg\}\Bigg) &\geq &c\,L\,\nu\Big(\big\{\omega_1 =[(\rho -\rho_c )L]\big\}\Big)\ .\\
& &\nonumber
\eea
\end{lemma}
\textbf{Proof.} The first statement is shown in \cite{vinogradov} Chapter~1, Corollary 1.1.1 to 1.1.3, the second in \cite{jeonetal00}, Theorem~2.2.\\
\\
\textbf{Proof of Theorem \ref{theo36}.}
The proof follows closely the one given in \cite{jeonetal00}, Theorem 2.2 and we only sketch the most important steps. Take $\rhobo\in A_i$ and $\mu=\overline\Mbo (\rhobo )$. To establish (\ref{lln}), using Theorem \ref{theo31}, it remains to show that for all $\epsilon >0$
\be
\lim_{L\to\infty} \tilde\nu_\mubo^{L,(i)} \Big(\big\{ M_L^i >\big(\rho_i -R_{c,i} (\rhobo )-\epsilon\big) L\big\}\Big) =1\ ,
\ee
where we use the shorthand $M_L^i (\feta )=\max_{x\in\Lambda_L} \eta_i (x)$. According to (\ref{condform}) the conditional measure of the inverse event is
\be\label{umr}
\tilde\nu_\mubo^{L,(i)} \big(\big\{ M_L^i \leq C_\epsilon L\big\}\big) = \frac{\nu_\mubo^{L,(i)} \Big(\Big\{ M_L^i \leq C_\epsilon L\, ,\,\Sigma_L^i =[\rho_i L]\Big\}\Big)}{\nu_\mubo^{L,(i)} \big(\{\Sigma_L^i =[\rho_i L]\}\big)}\ .
\ee
where $C_\epsilon :=\rho_i -R_{c,i} (\rhobo ) -\epsilon$. To prove the theorem we show that this expression vanishes for $L\to\infty$. We split the event
\be
\big\{ M_L^i \leq C_\epsilon L\big\} =\big\{ [L^\sigma ]\leq M_L^i \leq C_\epsilon L\big\}\cup\big\{ M_L^i <[L^\sigma ]\}
\ee
for some $\sigma\in (0,1)$ which is chosen below. A basic estimate in \cite{jeonetal00} shows that the probability of the second event vanishes for all $\sigma\in (0,1)$, whereas the first one is the crucial part. Analogous to the proof of Theorem \ref{theo31}, we single out one side that contains the maximum amount of particles. The maximum could be on any of the $L$ sites, so
\bea
\lefteqn{\nu_\mubo^{L,(i)} \Big(\Big\{ [L^\sigma ]\leq M_L^i \leq C_\epsilon L\, ,\,\Sigma_L^i =[\rho_i L]\Big\}\Big) =}\nonumber\\
& &\qquad =C\, L\sum_{[L^\sigma ]\leq k\leq C_\epsilon L} \nu_\mubo^{1,(i)} ( k)\,\nu_\mubo^{L-1,(i)} \big(\big\{\Sigma_{L-1}^i =[\rho_i L]-k\big\}\big)
\eea
where the constant $C\in\R$ accounts for over-counting some configurations where several sites have $k$ or more particles. It is derived in detail in \cite{jeonetal00}. Using monotonicity of $\nu_\mubo^{1,(i)}$ we obtain
\bea
\lefteqn{\nu_\mubo^{L,(i)} \Big(\Big\{ [L^\sigma ]\leq M_L^i \leq C_\epsilon L\, ,\,\Sigma_L^i =[\rho_i L]\Big\}\Big)\leq }\nonumber\\
& &\qquad\leq C\, L\,\nu_\mubo^{1,(i)} \big( [L^\sigma ]\big)\,\nu_\mubo^{L-1,(i)} \Big(\Big\{\Sigma_{L-1}^i \geq (R_{c,i} (\rhobo ) +\epsilon )(L-1)\Big\}\Big)\ .
\eea
With (\ref{ldresult}) the right-hand side is of order
\be
C\, L\,\nu_\mubo^{1,(i)} \big( [L^\sigma ]\big)\,\nu_\mubo^{1,(i)} \big(\big\{\eta_i \geq (R_{c,i} (\rhobo )+\epsilon )L\big\}\big) =\ord\big( L^{3-b(1+\sigma )}\big)\ .
\ee
On the other hand, the denominator of (\ref{umr}) is at least of order $L^{1-b}$ due to (\ref{ldresult}). Thus if we choose $\sigma\in (2/b ,1)$, (\ref{umr}) vanishes for $L\to\infty$ which finishes the proof.\hfill $\Box$\\

\section{Connection to zero-range processes}
The ensembles studied above arise naturally as stationary measures of zero-range processes showing a condensation transition which has recently attained much interest. Condensation transitions in zero-range processes with a single particle species have been studied in two cases: For site dependent jump rates of the particles \cite{benjaminietal96,evans96} the condensate is located at the slowest site. This case is closely related to Bose-Einstein condensation into the lowest energy level, which has been studied rigorously using large deviation techniques (see \cite{vandenbergetal88} and references therein). In this paper we consider the case of space homogeneous jump rates that induce an effective attraction between the particles \cite{evans00}. Such models have a number of direct applications, such as network dynamics or surface growth, and are particularly important in the study of phase separation in related exclusion models (see \cite{evansetal05} and references therein).

  \subsection{Definition}
The dynamics of a homogeneous zero-range process with two particle species on a finite, periodic lattice $\Lambda_L$ is defined by the generator
\be\label{generator}
\mathcal{L}f(\feta )=\sum_{i=1}^2 \sum_{x,y\in\Lambda_L} g_i \big(\feta (x)\big)\, p_i (y-x)\Big( f\big(\feta^{i,x\to y}\big) -f(\feta )\Big)\ .
\ee
Here $g_i (\feta (x))\in [0,\infty )$ is the rate at which site $x$ loses a particle of species $i$. It jumps to site $y$ according to an irreducible probability distribution $p_i$, and the resulting configuration is denoted by $\feta^{i,x\to y}$. We impose $g_i (\kbo )=0\ \Leftrightarrow\ k_i =0$ and thus the process is irreducible on each $X_{L,\Nbo}$ with fixed particle numbers. For finite lattices the generator is defined for all $f:X_L \to\R$, whereas on infinite lattices there are restrictions on the state space and the test functions $f$ \cite{andjel82}.

It has been shown \cite{evansetal03,stefan2} that for every positive weight $w:\N^2 \to (0,\infty )$ the zero-range process with rates
\be\label{rela}
g_1 (\kbo )=\frac{w(k_1 -1,k_2)}{w(k_1 ,k_2 )}\ ,\quad g_2 (\kbo )=\frac{w(k_1 ,k_2 -1)}{w(k_1 ,k_2 )}
\ee
has stationary product weight $w^L$ as defined in (\ref{weight}), independent of $p_i$. So independent of reversibility of the process, the canonical (\ref{ce}) and grand-canonical measures (\ref{gce}) are stationary and the results of Section 3 apply to the long-time behaviour of such processes. Thus our analysis on the static phase diagram, adopted from equilibrium statistical mechanics, applies also to non-equilibrium zero-range processes. On the other hand, dynamic quantities such as dynamic critical exponents or two-time correlation functions certainly depend on reversibility.

Note that (\ref{rela}) induces a relation between the rates $g_1$ and $g_2$, and not every two-species zero-range process has stationary product measures \cite{evansetal03,stefan2}. This is in contrast to single species systems, which always have product measures with $w(k)=\prod_{i=1}^k g(i)^{-1}$. In this case, $\partial D_\mu =\{\mu_c \}$ and $\partial D_\rho =\{\rho_c \}$ consist only of single points, and condensation has been directly related to the asymptotic behaviour of the jump rate \cite{evans00}. If $g(k)$ decays slower than $a+2/k$ as $k\to\infty$ for some $a\geq 0$, then $\rho_c <\infty$.

  \subsection{Generic examples}
In single-species zero-range processes, the asymptotic decay of the jump rate induces an effective attraction between the particles and results in convergence of $R(\mu )$ on $\partial D_\mu$. The same idea was used in \cite{evansetal03} to study an example of a two-species zero-range process with rates
\bea\label{exa1}
g_1 (\kbo )&=&\theta (k_1 )\left(\frac{k_1 (k_1 +2)}{(k_1 +1)^2}\right)^{k_2} \Big( 1+\frac{b}{k_1}\Big)\ ,\nonumber\\
g_2 (\kbo )&=&\theta (k_2 )\Big( 1+\frac{1}{k_1 {+}1}\Big)\ ,
\eea
where $\theta (0)=0$ and $\theta (k)=1$ for $k\geq 1$. The generic feature is that the rate of species $2$ particles depends only on the presence of species $1$ particles, and $g_1$ is then chosen to fulfill (\ref{rela}). This corresponds to the stationary weight (\ref{stawe})
\be
w(\kbo )=\frac{k_1 !}{(1+b)_{k_1}}\,\Big(\frac{k_1 +1}{k_1 +2}\Big)^{k_2}\ ,
\ee
which we already used as an example in Section 3. For $\mu_2 <0$ the grand-canonical partition function contains a geometric series and can be partially summed,
\bea\label{zexa1}
z(\mubo )&=&\sum_{k_1 =0}^\infty e^{\mu_1 k_1} \frac{k_1 !}{(1+b)_{k_1}} \sum_{k_2 =0}^\infty \Big(\frac{k_1 +1}{k_1 +2}\Big)^{k_2} e^{\mu_2 k_2} =\nonumber\\
&=&\sum_{k_1 =0}^\infty e^{\mu_1 k_1}\frac{2+k_1}{(1-e^{\mu_2})(k_1 +1)+1}\frac{k_1 !}{(1+b)_{k_1}} \ .
\eea
So $\dom z$ and thus $\Di_\mu$ is a rectangle with $\Di_\mu =\big\{\mubo\,\big|\,\mu_1 ,\mu_2 <0\big\}$. The parts of the boundary that belong to $D_\mu$ depend on the parameter $b$, resulting in different phase diagrams, as illustrated in Figures \ref{fig1} and \ref{fig2} in Section 3. For $\mu_1 =0$ the factor
\be\label{pochhammer}
\frac{k_1 !}{(1+b)_{k_1}} =k_1 !\Big/\prod_{i=0}^{k_1 -1} (i+b)\sim k_1^{-b}\quad\mbox{as }k_1 \to\infty\ ,
\ee
in (\ref{zexa1}) determines the convergence properties of $\Rbo (\mubo )$ on $\partial D_\mu$:
\bea
b>3\quad &\Rightarrow &\quad D_\mu =\big\{\mubo\,\big|\,\mu_1 ,\mu_2 \leq 0\big\}\ ,\nonumber\\
3\geq b>2\quad &\Rightarrow &\quad D_\mu =\big\{\mubo\,\big|\,\mu_1 ,\mu_2 \leq 0\big\}\setminus\big\{ (0,0)\big\}\ ,\nonumber\\
b\leq 2\quad &\Rightarrow &\quad D_\mu =\big\{\mubo\,\big|\,\mu_1 <0,\mu_2 \leq 0\big\}\ .
\eea

The boundary between the phase regions $D_\rho$ and $A_2 \setminus A_1$ for $\mu_2 =0$ can be calculated explicitly as $R_2 (\mu_1 ,0)=1+R_1 (\mu_1 ,0)$, $\mu_1 <0$, whereas the other boundary for $\mu_1 =0$ is only given implicitly by $R_i (0,\mu_2 )=\partial_{\mu_i} z(0,\mu_2 )$ (see \cite{evansetal03} for more details). On top of the stationary phase diagram discussed here, the relaxation dynamics of this zero-range process shows an interesting coarsening phenomenon, which has been analyzed in \cite{stefan3}.

In the following we consider two other examples which have not been studied before. The first one, demonstrating that $D_\mu$ does not have to be a rectangle, is
\be\label{exa2}
g_1 (\kbo )=\theta (k_1 )\left(\frac{k_1}{1+k_1} \right)^{k_2} (1+b/k_1)\ ,\quad g_2 (\kbo )=\frac{k_2}{1+k_1}\ .
\ee
Here $g_2 \propto k_2$ so particles of the second species move independently but are slowed down by the presence of species $1$ particles, and again $g_1$ is chosen to fulfill (\ref{rela}). The corresponding stationary weight is
\be\label{wexa2}
w(\kbo )=\frac{k_1 !}{(1+b)_{k_1}}\,\frac{(k_1 +1)^{k_2}}{k_2 !}\ .
\ee
Also in this model the partition function can be partially summed and written as a hypergeometric function
\be\label{wexa3}
z(\mubo )=e^{e^{\mu_2}}\sum_{k_1 =0}^\infty \frac{k_1 !}{(1+b)_{k_1}}\, e^{(e^{\mu_2}+\mu_1 )k_1} =e^{e^{\mu_2}} {_2 F_1 }\big( 1,1;1+b;e^{e^{\mu_2}+\mu_1 }\big)\ .
\ee
So $D_\mu =\big\{\mubo\big| e^{\mu_2} +\mu_1 \leq 0\big\}$ is a closed set for $b>2$ due to (\ref{pochhammer}). Since $\partial D_\mu$ is curved, the only non-empty condensing phase region is $A_1 \cap A_2$. The resulting phase diagram is shown in Figure \ref{fig3}. As in Figure \ref{fig1}, the dashed lines point in the normal directions of $\partial D_\mu$ and in the density plane they determine the background density $\Rbo_c (\rhobo )$. In the above example with rates (\ref{exa1}) these lines are actually uniquely determined by the phase boundaries alone (see Figure \ref{fig1}, right), whereas for this new example they have to be fixed via the normal vectors $n_\mubo \parallel (1,e^{\mu_2} )$ of $\partial D_\mu$. Thus Theorem \ref{theo33} implies that the background densities fulfill
\be
\frac{\rho_2 -R_{c,2} (\rhobo)}{\rho_1 -R_{c,1} (\rhobo)} =\frac{e^{\overline\Mbo_2 (\rhobo )}}{1}\quad\mbox{for all }\rhobo\in A_1 \cap A_2\ .
\ee
Using (\ref{wexa2}) and (\ref{wexa3}) it is easy to see that $R_2 (\mubo )=\big( 1+R_1 (\mubo )\big)\, e^{\mu_2}$ for all $\mubo\in D_\mu$. Furthermore
\be
R_1 (\mubo )=\partial_{\mu_1} z(\mubo )=\frac1{b-2}\quad\mbox{for all }\mubo\in\partial D_\mu \ ,
\ee
using standard expansions of the hypergeometric function $_2 F_1$ which are summarized e.g.\ in \cite{stefan}. Taken together, this implies that for $\rhobo\in A_1 \cap A_2$
\be
R_{c,1} (\rhobo )=\frac{1}{b-2}\ ,\quad R_{c,2} (\rhobo )=\Big( 1+\frac{1}{b-2}\Big)\frac{\rho_2}{1+\rho_1}\ .
\ee
By coincidence, the lines $\big\{\rhobo\,\big|\,\Rbo_c (\rhobo )=(\frac{1}{b-2},\rho_{c,2})\big\}$ converge in the point $(-1 ,0)$ for all $\rho_{c,2} >0$, as is shown in Figure \ref{fig3} on the right.
\bef
\includegraphics[width=0.45\textwidth]{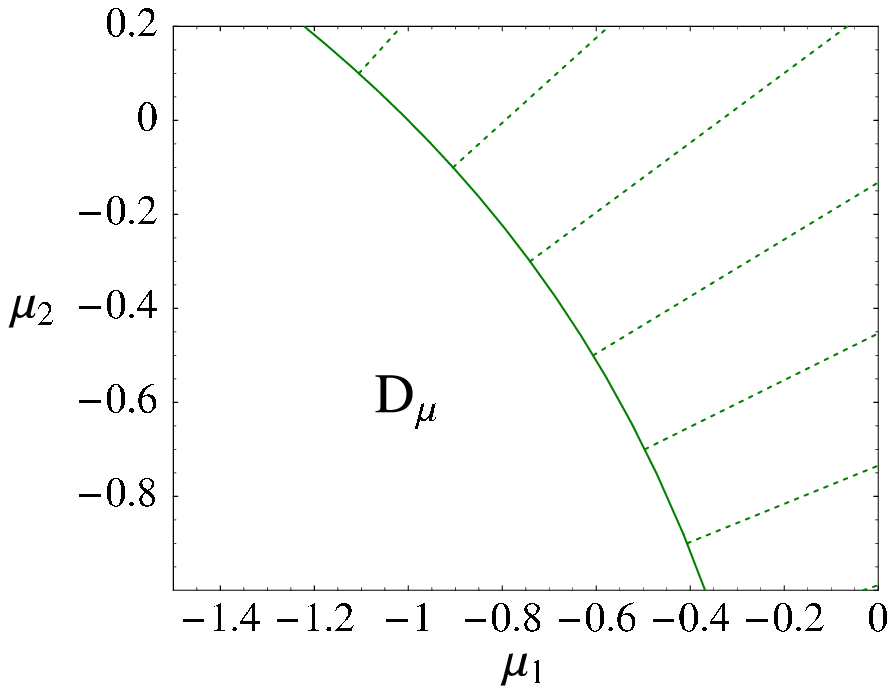}\hfill\includegraphics[width=0.45\textwidth]{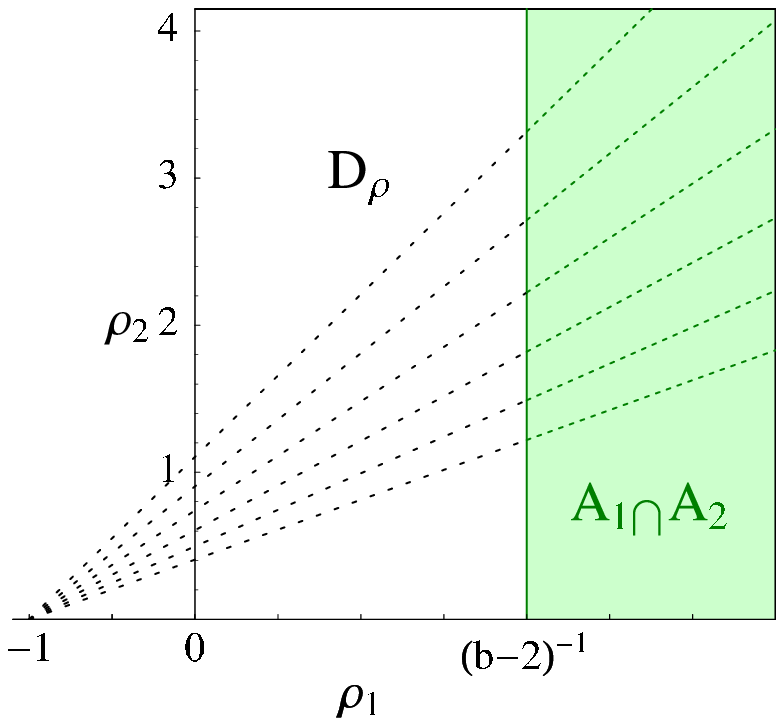}
\caption{\label{fig3}$D_\mu$ and phase diagram for the zero-range process with rates (\ref{exa2}) for $b=4$.  Dashed lines on the left denote the normal directions to $\partial D_\mu$ and on the right they determine the background density $\Rbo_c (\rhobo )$.}
\enf

If $w$ is composed of several parts, one can produce various kinds of phase diagrams. For example if we add $w(k_2 ,k_1 )$ to (\ref{wexa2}) we get the symmetrized version
\be\label{exa3}
w(\kbo )=\frac{k_1 !}{(1+b)_{k_1}}\,\frac{(k_1 +1)^{k_2}}{k_2 !}+\frac{k_2 !}{(1+b)_{k_2}}\,\frac{(k_2 +1)^{k_1}}{k_1 !}\ .
\ee
The domain is then given by the intersection of $D_\mu$ from (\ref{wexa2}) with its symmetric counterpart, i.e.
\be
D_\mu =\big\{\mubo\big| e^{\mu_2} +\mu_1 ,e^{\mu_1} +\mu_2 \leq 0\big\}\ .
\ee
This is illustrated in Figure \ref{fig4} together with the phase diagram, where phase region $A_1 \cap A_2$ now shows two different kinds of behaviour of the function $\Rbo_c (\rhobo )$.
\bef
\includegraphics[width=0.45\textwidth]{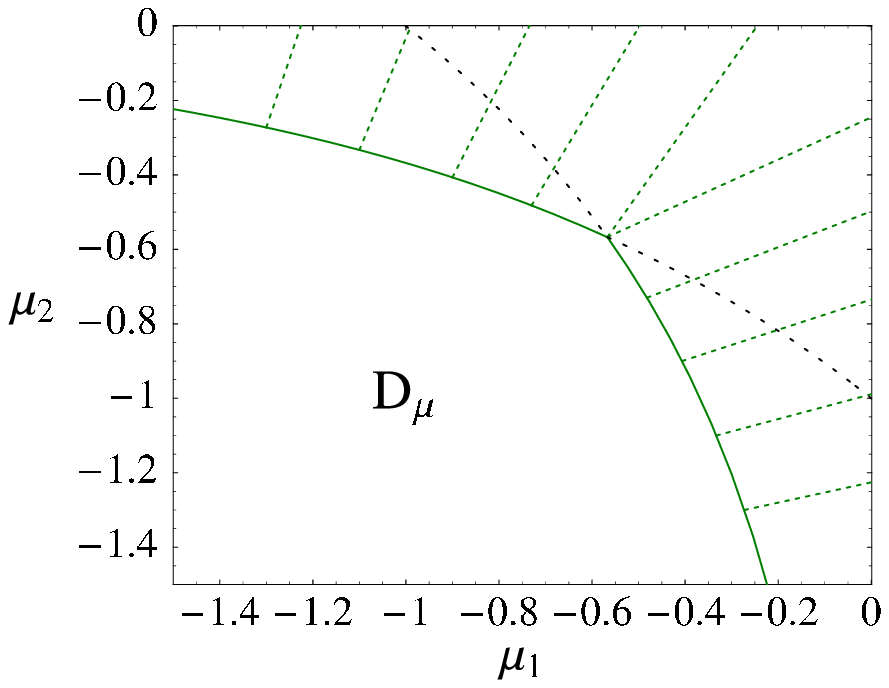}\hfill\includegraphics[width=0.45\textwidth]{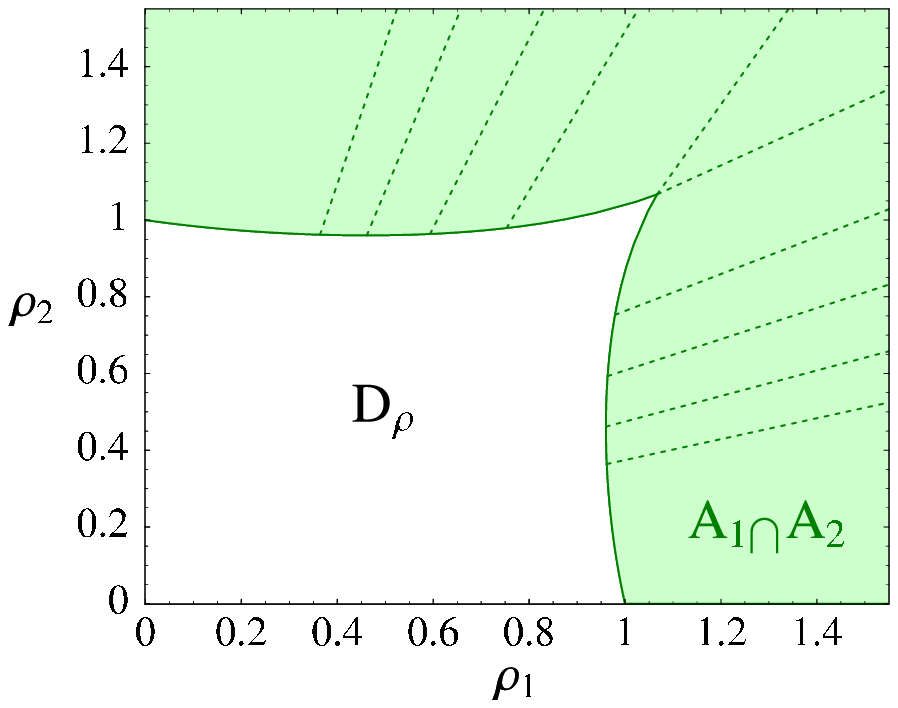}
\caption{\label{fig4}$D_\mu$ and phase diagram for the stationary weight (\ref{exa3}) for $b=4$, analogous to Figure \ref{fig3}.}
\enf

In this way one can find zero-range processes exhibiting all kinds of phase diagrams. However, these models are often artificial since the jump rates are very complicated due to the constraint (\ref{rela}), in particular for the last example. On the other hand, simple rates may lead to zero-range processes for which the stationary distribution is unknown and not of product form. For such models, a recent study revealed the possibility of a discontinuous condensation transition \cite{godreche06}.

  \subsection{Further remarks}
In section 4.3 we noted that the regularity condition (\ref{cond2}) on the exponential tail of $w$ can be relaxed considerably. However, for the results of Section 3 to hold one has to assume some regularity of $w$. Consider for example a single-species zero-range process with stationary weight
\be\label{counterw}
w(k)=\left\{\bacl k^{-3} \ &,\ k\in I\\ 2^{-k}\ &,\ k\not\in I\ea\right.\ ,
\ee
for some set $I=\{ i_1 ,i_2 ,\ldots\}\subset\N$. If $|I|=\infty$ then $D_\mu =(-\infty ,0]$. The proof of Theorem \ref{theo33} works as long as $\sup_{j\in\N} (i_{j+1}-i_j ) <\infty$. This is not a purely technical condition, because if it is violated, e.g.\ for $i_j =2^j$, one does not expect the condensate to be stable, since it cannot fluctuate in size. Such a stationary weight leads to jump rates with exponentially growing variation as $k\to\infty$, so Monte-Carlo simulations for such processes are not feasible. But the behaviour of such irregular processes is in general only of limited interest.

Following the results in Section 3.3, in phase region $A_1 \cap A_2$ we expect the condensate of each species to concentrate on a single lattice site. Moreover, in the examples in Section 5.2 these two condensates are expected to be on the same lattice site, since $g_1$ is a decreasing function of $k_2$ and vice versa, inducing an effective attraction between the condensates. Indeed this is what is found in simulations \cite{stefan3}. If both species are independent the stationary weight factorizes, i.e.\ $w(\kbo )=w_1 (k_1 )\, w_2 (k_2 )$, the condensates do not interact and both have independent random positions. On the other hand, if $g_1$ is increasing in $k_2$ and vice versa, the condensates repel each other and are not found on the same site. In general, whenever the species are coupled, the presence of a condensate of one species influences the distribution of the other species on that site, also if only one species condenses. This effect is important for the analysis of the coarsening behaviour for two species systems and is studied heuristically in \cite{stefan3}.

All results of Section 3 only address the stationary distribution of zero-range processes. Apart from studying the coarsening dynamics, the ergodic behaviour of the system on an infinite lattice $\Lambda$ is a dynamical question which is expected to be closely related to the stationary results. Starting with a homogeneous distribution $\mu_\rhobo (0)$ with density $\rhobo =\big\langle\feta (x)\big\rangle_{\mu_\rhobo (0)}$, $x\in\Lambda$ at time $t=0$, we expect as $t\to\infty$
\be\label{dyna}
\mu_\rhobo (t)=\mu_\rhobo (0)\, e^{\mathcal{L}t}\stackrel{d}{\longrightarrow}\nu_{\overline\Mbo (\rhobo )}\ ,\quad i.e.\quad\langle f\rangle_{\mu_\rhobo (t)} \to\langle f\rangle_{\nu_{\overline\Mbo (\rhobo )}}
\ee
for bounded cylinder test functions $f\in C_b (X,\R )$. Although $\overline\Mbo$ (see Proposition \ref{theo22}) is the same function as for the stationary results, (\ref{dyna}) is a statement about the dynamics of a zero-range process (\ref{generator}) and requires a completely different analysis. Such ergodic results exist for attractive single species systems with non-decreasing jump rates $g(k)$, where one can use coupling techniques. These are not applicable in case of condensation, since then even the single-species process is not attractive, and we are not aware of other results in this direction.\\

\section{Summary}
In this paper we adapt the theory of the equivalence of ensembles to study phase separation in particle systems with unbounded local state space. Our results cover condensation transitions in (multi-species) zero-range processes which are currently of particular interest, and are the main motivation for this study. We use the method of specific relative entropy, previously applied to systems with bounded Hamiltonians, which in our case involves large deviations and multivariate local limit theorems of subexponential distributions. We derive the phase diagram for the condensation and explain its connection to the mode of convergence in the equivalence of ensembles, generalizing previous results for the non-condensing case. Condensation is shown to be a continuous phase transition, where the mechanism is different from systems with bounded Hamiltonian, and can be characterized by convergence properties of the Gibbs free energy on the boundary of its domain of definition. The analysis also involves interesting properties of the boundary behaviour of multivariate power series.

For simplicity of presentation we formulate our results not in the most general setting, but focus on a particular case which captures the basic novelties of the paper with respect to previous work, and is closely related to the main application to zero-range processes. A generalization to any number of particle species, each having arbitrary discrete state space is straightforward, as long as the stationary measures are of product form. Since the method of specific relative entropy only makes use of permutation invariance, an extension to non-product measures along the lines of \cite{lewisetal95} is possible, but requires a substantial amount of work. Single-species processes leading to measures with nearest-neighbour Hamiltonians have recently been investigated non-rigorously \cite{evansetal06}. The result for the equivalence of ensembles is expected to be the same, but the structure of the condensate should be different due to spatial correlations. Another open point is a rigorous result on the structure of the condensate for more than one species, which involves large deviations for multivariate subexponential distributions.\\

\section*{Acknowledgments}
The author would like to thank H.-O.~Georgii and H.~Spohn for careful reading of the manuscript and very useful discussions, A.~Zaigraev and C.-E.~Pfister for valuable advice on previous results. The author is also grateful for the kind hospitality of the Isaac Newton Institute, Cambridge and for helpful discussions with colleagues at the programme 'Principles of the Dynamics of Non-Equilibrium Systems', M.R.~Evans, C.~Godr\`eche and T.~Hanney.

\appendix

\section{Results from convex analysis}
In the following we summarize a few results from convex analysis which are used in the paper, taken from \cite{rockafellar}.

For a function $f:\R^n \to\overline\R =\R\cup\{ -\infty ,\infty\}$ we denote by
\be
\dom\, f=\big\{ x\in\R^n \,\big|\, |f(x)| <\infty\big\}
\ee
the \textit{domain} of $f$. $f$ is called \textit{proper} if $\dom\, f\neq\emptyset$ and $f(x)>-\infty$ for all $x\in\R^n$. For example any function $f:D\to\R$ defined on some $D\subset\R^n$ can be extended to a proper function by setting $f=\infty$ on $\R^n \setminus D$. To simplify some points we will often concentrate on proper functions $f:\R^n \to\R\cup\{\infty\}$ in the following. A proper function $f$ is \textit{convex} if for all $x,y\in\R^n$ and $\tau\in (0,1)$
\be\label{conv}
f\big( (1-\tau )x +\tau y\big)\leq(1-\tau )f(x)+\tau f(y)\ .
\ee
Note that for $x\not\in\dom\, f$ or $y\not\in\dom\, f$ (\ref{conv}) holds trivially, and in particular it implies that $\dom f$ has to be a convex set. $f$ is \textit{strictly convex} if (\ref{conv}) holds with strict inequality. We denote by
\be
{\rm argmin} f=\big\{ x\in\dom\, f\,\big|\, f(x)=\inf_{x\in\dom f} f(x)\big\}
\ee
the set of minimizers of $f$.
\begin{theorem}\label{conmin}
Let $f:\R^n \to\overline\R$ be convex. Then ${\rm argmin} f$ is a convex subset of $\R^n$. If $f$ has a local minimum in $x\in\dom\, f$ then $x\in {\rm argmin} f$, i.e.\ every local minimum is a global minimum. If $f$ is strictly convex ${\rm argmin} f$ is either a singleton or empty.
\end{theorem}
\textbf{Proof.} See \cite{rockafellar}, Theorem 2.6.\\
\\
$f:\R^n \to\overline\R$ is called \textit{lower semicontinuous at $x$} if
\be
\liminf_{y\to x} f(y):=\lim_{\epsilon\searrow 0} \big(\inf_{y\in B(x,\epsilon )} f(y)\big) =f(x)\ ,
\ee
and \textit{lower semicontinuous} if this holds for every $x\in\R^n$. We denote by
\be
\lev_{\leq\alpha }f:=\{ x\in\R^n \,\big|\, f(x)\leq\alpha\}\quad\mbox{and}\quad\lev_{=\alpha }f:=\{ x\in\R^n \,\big|\, f(x)=\alpha\}
\ee
level sets and level lines of a proper function $f$.
\begin{theorem}\label{levels}
$f$ is lower semicontinuous if and only if the level sets $\lev_{\leq\alpha }f$ are closed in $\R^n$ for all $\alpha\in\R$. If $f$ is convex, then the level sets $\lev_{\leq\alpha }f$ are convex.
\end{theorem}
\textbf{Proof.} See \cite{rockafellar}, Theorem 1.6. and Proposition 2.7.\\
\\
For a proper convex function, the \textit{subgradient} at $x\in\dom\, f$ is given by
\be\label{subgr}
\delta f(x):=\big\{ v\in\R^n\,\big|\, f(y)\geq f(x)+v\cdot (y-x)\mbox{ for all }y\in\dom\, f\big\}\ .
\ee
If $f$ is differentiable in $x$, then $\delta f(x)=\big\{\grad f(x)\big\}$. There is a more general definition of subgradients for non-convex functions (see \cite{rockafellar}, Definition 8.3), which we omit since we do not make use of it. It is consistent with (\ref{subgr}) for convex functions as proved in \cite{rockafellar}, Proposition 8.12.\\
For any function $f:\R^n \to\overline\R$, the \textit{convex conjugate} function $f^* :\R^n \to\overline\R$ is given by
\be\label{legend}
f^* (v)=\sup_{x\in\R^n} \big( v\cdot x-f(x)\big)\ ,
\ee
and the mapping $f\mapsto f^*$ is called the \textit{Legendre-Fenchel transform}. 
\begin{theorem}\label{duality}
Let $f:\R^n \to\overline\R$ be a proper convex function. Then $f^*$ is proper, convex and lower semicontinuous. If $f$ is lower semicontinuous in $x\in\dom\, f$, then
\be
v\in\delta f(x)\quad\Leftrightarrow\quad x\in\delta f^* (v)\quad\Leftrightarrow\quad f(x)+f^* (v)=v\cdot x\ .
\ee
\end{theorem}
\textbf{Proof.} See \cite{rockafellar}, Theorem 11.1 and Proposition 11.3.\\

\section{Construction of ensembles on a common space}
To formulate the convergence result (\ref{corr1}) we need to define the sequences of canonical measures and the grand-canonical product measure on a common measurable space. 

Take $\Lambda_L \subsetneq\Lambda_{L+1}$ and $\Lambda_L \uparrow\Lambda$ as $L\to\infty$, i.e. $\Lambda_L \subsetneq\Lambda$ for all $L$ and for all $x\in\Lambda$ there exists $L\geq 1$ such that $x\in\Lambda_L$. Set $X=\N^\Lambda$ and let $\mathcal{A}$ be the $\sigma$-algebra induced by the product topology on $X$, which is generated by the set of all cylinder configurations. We can identify $X_L$ by the set of cylinder configurations on $\Lambda_L$, which are
\be
\feta =\big\{ \fzeta\in X\,\big|\,\fzeta (x)=\feta (x)\mbox{ for all }x\in\Lambda_L \big\}\ ,
\ee
so that $\mathcal{A}=\sigma (X_1 ,X_2 ,\ldots )$. The family $\nu_\mubo^L$ directly extends to a product measure $\nu_\mubo$ on $(X,\mathcal{A} )$ such that $\nu_\mubo (\feta )=\nu_\mubo^L (\feta )$ for all $\feta\in X_L$, $L\geq 1$. On the other hand, $\pi_{L,\Nbo}$ is a measure on $\big( X,\sigma (X_L )\big)$ where $\sigma (X_L )$ is the $\sigma$-algebra generated by $X_L$, i.e. the smallest $\sigma$-algebra that contains all cylinder configurations $\feta\in X_L$. Then $\pi_{L,\Nbo}$ is also a measure on $\big( X,\sigma (X_k )\big)$ for all $k\leq L$, since $\Lambda_L \subsetneq\Lambda_{L+1}$ implies $\sigma (X_L )\subsetneq\sigma (X_{L+1} )$. For a cylinder function $f:X\to\R$ there exists a fixed $n\in\N$, such that $f$ is $\sigma (X_n )$-measurable, i.e. $f$ depends only on coordinates in $\Lambda_n$. Then we take $\big( X,\sigma (X_n )\big)$ as the measurable space on which $\pi_{L,\Nbo}$ and $\nu_\mubo^L$ are defined for $L\geq n$. This construction is sufficient to make sense of (\ref{corr1}), since it only addresses the limit $L\to\infty$.

\bibliography{zr2cond}
\bibliographystyle{unsrtnat}

\end{document}